\documentclass[aps, prd, tightenlines, notitlepage, superscriptaddress, twocolumn, nofootinbib, preprintnumbers, floatfix, showkeys,10pt]{revtex4-2}

\input{header}

\begin{document}

\title{Muon Tridents at FASER and NA64$\mu$: \\Motivation, Predictions, and Prospects for Precision Measurements}

\author{Reinaldo Francener\,\orcidlink{0000-0001-7176-388X}}
\email{reinaldofrancener@gmail.com}
\affiliation{Institute of Physics and Mathematics, Federal University of Pelotas (UFPel), \\ Postal Code 354,  96010-900, Pelotas, RS, Brazil.}
\author{Victor P. Gon\c{c}alves\,\orcidlink{0000-0003-4943-9973}}
\email{barros@ufpel.edu.br}
\affiliation{Institute of Physics and Mathematics, Federal University of Pelotas (UFPel), \\ Postal Code 354,  96010-900, Pelotas, RS, Brazil.}

\author{Felix Kling\,\orcidlink{0000-0002-3100-6144}}%
\email{fkling@uni-bonn.de}
\affiliation{Universit\"at Bonn, Regina-Pacis-Weg 3, D-53113 Bonn, Germany}

\author{Gabriel Rabelo-Soares\,\orcidlink{0000-0003-3843-4747}}
\email{rabelosoares21@gmail.com}
\affiliation{Instituto de Física Gleb Wataghin - Universidade Estadual de Campinas (UNICAMP), \\ 13083-859, Campinas, SP, Brazil. }

\begin{abstract}
The intense flux of high-energy muons produced in the far-forward region of $pp$ collisions at the LHC provides a unique opportunity to study rare electromagnetic interactions at TeV energies. In this work, we investigate the muon trident scattering in muon-nucleus interactions at the FASER experiment, which is expected to have recorded more than $10^5$ such events. The theoretical description of the process is reviewed and improved. Predictions for the total event rates and differential distributions expected during Run~3 of the LHC are presented, taking into account realistic detector acceptance and the main background contributions. The dependence of the predictions on the treatment of the nuclear form factors and the impact of higher-order QED effects such as Coulomb corrections are investigated. Our results indicate that the muon trident process can be measured for the first time at TeV energies using FASER's spectrometer, with a projected precision on the cross-section at the percent level. We also extended our analysis to the NA64$\mu$ experiment and showed that, together with FASER, these experiments can probe the energy dependence of the muon trident cross-section over a broad unexplored energy range.
\end{abstract}

\maketitle

\section{Introduction} The electromagnetic production of charged muon pairs by an incident muon in the Coulomb field of a heavy nucleus, usually denoted as the muon trident process, is a basic prediction of Quantum Electrodynamics (QED) and its determination is fundamental to determine the electromagnetic energy loss of high-energy muons propagating through matter, needed for the analysis of data produced by Neutrino and Cosmic Ray Observatories. Despite more than half a century of experimental studies of this QED process, surprisingly few such events were recorded, and significant discrepancies remain between different theoretical predictions available in the literature.

The first observations of this process was reported in the late 1960s using atmospheric muons~\cite{Morris1968, Barton1970}, which have motivated its search in laboratory. Measurements of the muon trident cross-section were published in 1971~\cite{Russell:1971mf} and 1981~\cite{Lebritton:1979gd}, both using a 10.5~GeV muon beam at the Alternating Gradient Synchrotron (AGS) at BNL. In these two measurements, $75 \pm 10$ and $137 \pm 14$ candidates were observed, respectively, and used to constrain the magnitude of the associated cross-section and to confirm that muons obey the Fermi-Dirac statistics. In the following decades, additional experiments using atmospheric muons reported muon trident candidate events at few GeV energies. These include hints of the process in MACRO~\cite{MACRO:1999qxo, Battistoni:409814}, seven candidate events in BARS~\cite{Anikeev:2005td}, and two candidate events in ALEPH~\cite{Maciuc:2006xb}. However, the limited number of observed events and the difficulty in reconstructing the energies of the incoming and outgoing muons have prevented a precise measurement of the cross-section. On the theoretical side, several authors have estimated the muon trident cross-section, with its predictions differing by more than one order of magnitude~\cite{Henry:1967jm, Tannenbaum:1968zz, Bugaev1970, Barger:1979eg, Ganapathi:1980wh, Kelner:1998mh, Kelner:2000va,Das:2026eyy}. Such differences arise from the distinct approximations assumed in the calculations, e.g., the use (or not) of the equivalent photon approximation (EPA) \cite{Budnev:1975poe} and the treatment of the nuclear form factor. Both the limited statistics of previous more than 40-year-old measurements on the experimental side and the discrepancies between predictions on the theoretical side are strong motivations to revisit this process and investigate if it can be measured using the current and forthcoming experiments. 

Progress in understanding the muon trident process $\mu^{\pm} A \rightarrow \mu^{\pm} \mu^+ \mu^- A$ over the past few decades was hindered by the lack of experimental capabilities to measure its final state. However, recently it became clear that $pp$ collisions at the LHC generate a flux of muons that can reach the far-forward LHC detectors, FASER~\cite{FASER:2022hcn} and SND@LHC~\cite{SNDLHC:2022ihg}, which were originally designed to search for decays of long-lived particles predicted by models of new physics~\cite{Feng:2017uoz} as well as to detect and study high-energy neutrinos produced at the LHC~\cite{FASER:2019dxq}. Measurements of this flux have been performed by both collaborations~\cite{FASER:2021mtu, SNDLHC:2023mib, SNDLHC:2026mip}, which motivated the investigation of the physics opportunities offered by this muon flux and a renewed interest in the muon trident process. In particular, Refs.~\cite{Francener:2025pnr, Francener:2025tyh,Hernandez-Sainz:2026rxc} demonstrated that high-energy muon deep-inelastic scattering (DIS) interactions inside neutrino detectors provide a powerful probe of proton and nuclear structure. In addition, Refs.~\cite{Ariga:2023fjg, Batell:2024cdl, MammenAbraham:2025gai} investigated the sensitivity of these muons to beyond-the-Standard-Model (BSM) signatures involving muon-philic particles. Regarding the muon trident process, Ref.~\cite{Francener:2025wzh} showed that FASER's emulsion detector can be used to measure muon-induced lepton trident production for all charged-lepton final states during Run~3 of the LHC, and the SND@LHC collaboration has recently reported the first candidate events for muon trident events\footnote{Note that the SND@LHC detector does not contain a magnetized spectrometer. Therefore, no muon momentum information is available for these events.} occurring in the rock upstream of the detector~\cite{SNDLHC:2024bzp, Guler:2026mrt}. 

\begin{figure}[t]
	\centering
	\begin{tabular}{c}
\includegraphics[width=0.47\textwidth]{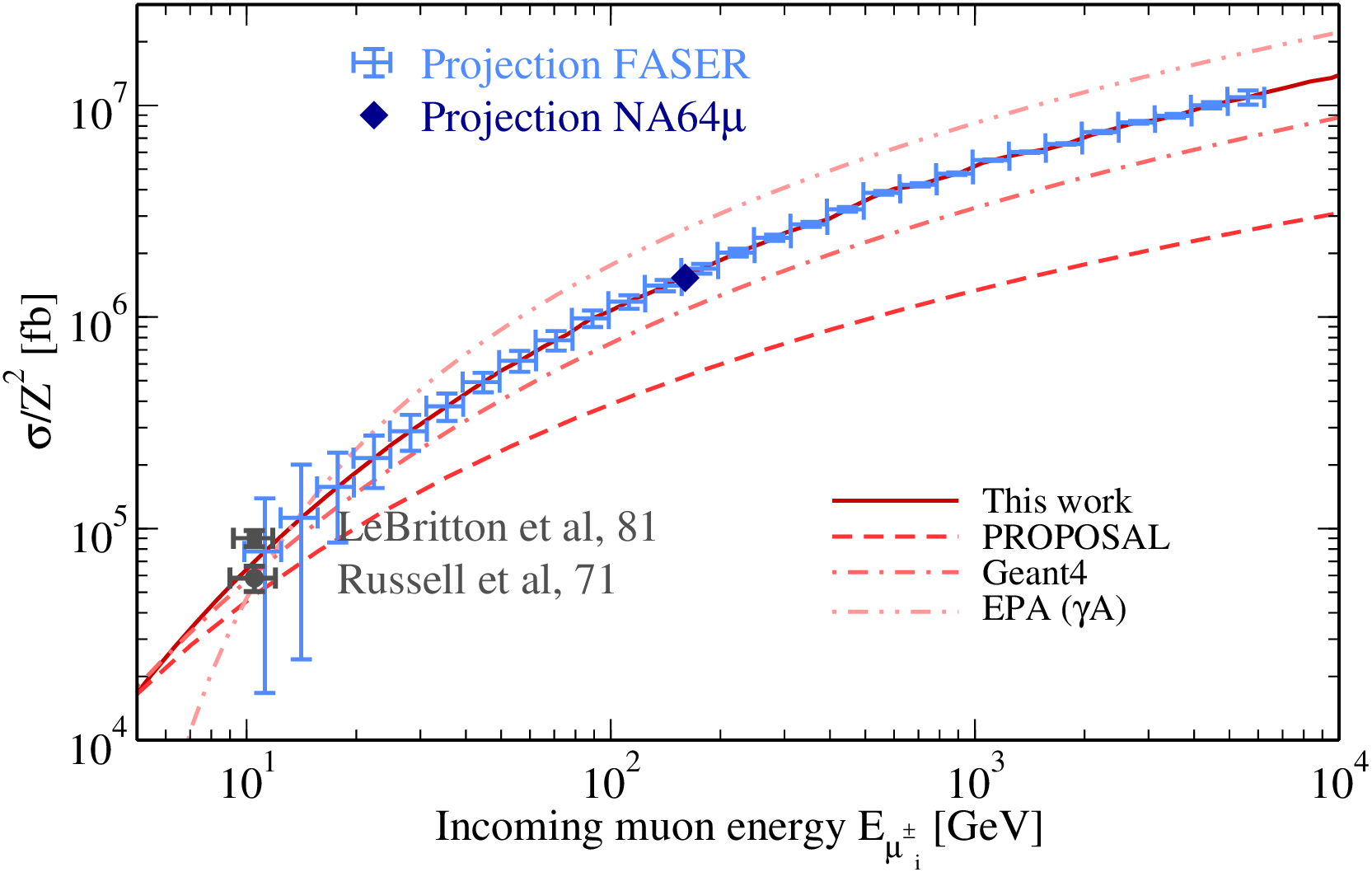} 
	\end{tabular}
\caption{Predicted sensitivity for the muon trident cross-section as a function of incident muon energy at FASER and NA64$\mu$. Cross-section is normalized by the atomic number to minimize differences for tungsten and lead targets. We included also previous measurements with 10.5~GeV muons from Refs.~\cite{Russell:1971mf, Lebritton:1979gd}. }
\label{fig:projected_sensitivity}
\end{figure}

In this work, we will perform a detailed investigation of the potential of the FASER experiment to measure muon trident process. In contrast to Ref.~\cite{Francener:2025wzh}, we will consider the measurement using only the electronic detector components, in particular FASER's scintillator system and spectrometer. This will allow for a precise reconstruction of the final-state muon momenta up to TeV energies and the determination of muon trident rate differentially, with high statistics, over a large kinematic range. Moreover, predictions for the NA64$\mu$ experiment will also be presented. The impact of the different theoretical assumptions on the final results will be investigated. In particular, we will analyze the dependence on the description of the nuclear form factor, estimate the impact of the interference term associated with the presence of two identical muons in the final state~\cite{Tannenbaum:1968zz, Russell:1971mf, Lebritton:1979gd} and calculate, for the first time, the contribution of Coulomb corrections~\cite{Davies:1954zz, Baltz:2009fs, Jentschura:2009mb, Li:2023yjt} for the muon trident process. Finally, the sensitivity of the muon trident process for electroweak corrections and to New Physics will also be discussed.

Our main results are summarized in Fig.~\ref{fig:projected_sensitivity}, where we present the predicted sensitivity for the muon trident cross-section as a function of incident muon energy at FASER and NA64$\mu$. The cross-section has been normalized by the atomic number in order to minimize differences for tungsten (FASER) and lead (NA64$\mu$) targets. Moreover, previous measurements with 10.5~GeV muons from Refs.~\cite{Russell:1971mf,Lebritton:1979gd} are also presented. As we will demonstrate below, the full calculations performed in our analysis, reproduce the predictions obtained using \texttt{Geant4}~\cite{GEANT4:2002zbu, Yajaman2022} and \texttt{Proposal}~\cite{Koehne:2013gpa, Dunsch:2018nsc} software packages at small incoming muon energies and reproduce the current data, but imply larger cross-sections at higher energies. In addition, we will show that the muon trident cross-section can be measured, for the first time, over a range of incoming muon energies from a few tens of GeV to several TeV, with a projected percent-level precision based on more than $8.5\times 10^4$ events during Run~3 of the LHC.

This paper is organized as follows. In the next section, we review the theoretical description of the muon trident process and discuss the main ingredients present in the calculation of the associated cross-section, as well the approximations usually assumed in the literature. The experimental setups for the FASER and NA64$\mu$ experiments assumed in our analysis are presented in Section~\ref{sec:experimental}. Moreover, the potential backgrounds for the measurement of the muon trident process will be discussed. Our results for the total cross-sections and differential distributions are presented in Section~\ref{sec:results}. Finally, in Section~\ref{sec:sum}, we summarize our main results and conclusions.

\section{Cross-Section Modeling}
\label{sec:cross_section_modeling}

A pair of leptons can be produced in muon-nucleus interactions through coherent, incoherent and deep inelastic scattering processes. These distinct processes have different dependencies on the nuclear charge $Z$ and are characterized by distinct final states. On the one hand, coherent processes are proportional to $Z^2$, with the nucleus remaining intact after the interaction. On the other hand, in incoherent and deep inelastic scattering processes the nucleus breakup, with the interaction occurring between the muon and the nucleons and partons inside the nucleus, respectively, and the cross-section being proportional to $Z$. Deep inelastic contributions are additionally suppressed by the partonic structure functions in the kinematic regime considered in this work. In principle, these contributions can be experimentally separated if the interaction vertex is reconstructed, then the coherent process will be characterized by an environment without hadronic activity. 

\begin{figure}[t]
\centering
\begin{tabular}{cc}
\includegraphics[width=0.23\textwidth]{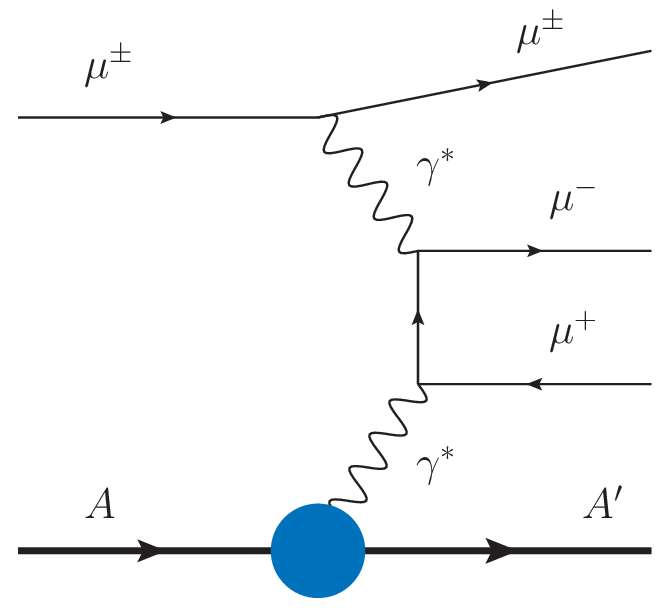} 
\includegraphics[width=0.23\textwidth]{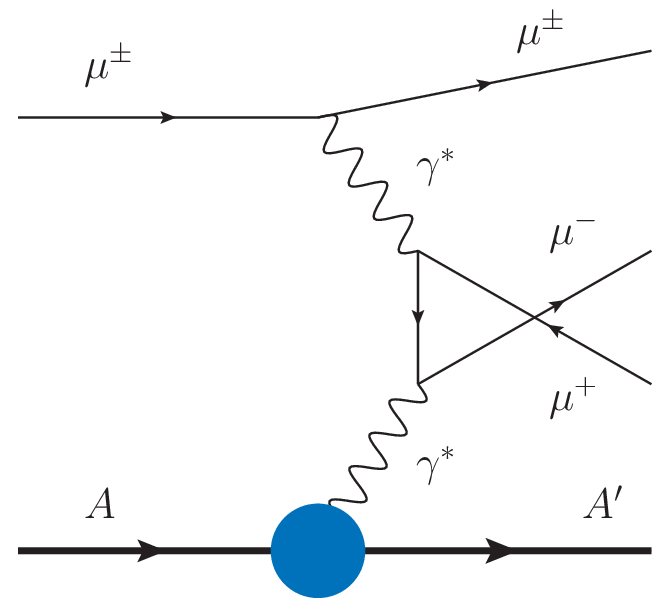} \\
\includegraphics[width=0.23\textwidth]{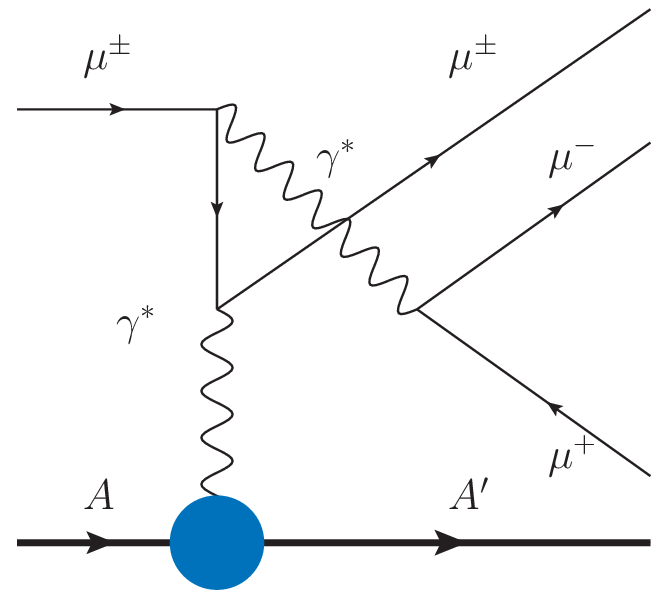} 
\includegraphics[width=0.23\textwidth]{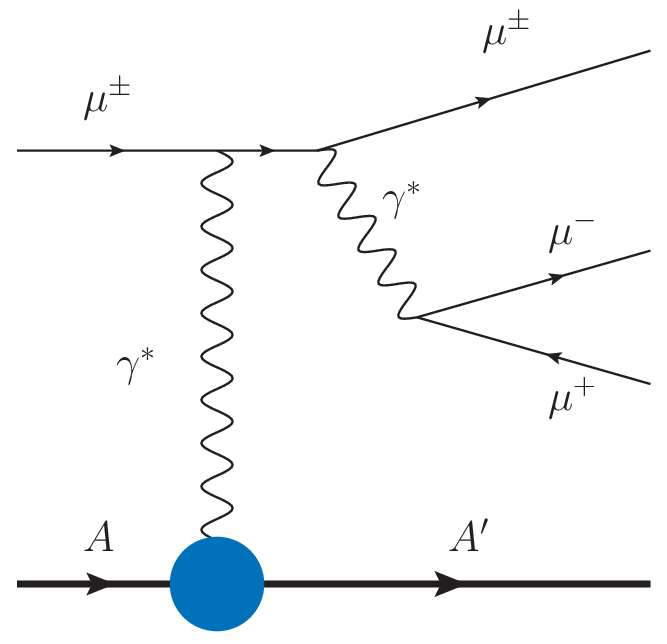} 
\end{tabular}
\caption{ Leading order diagrams for the muon trident process in the Coulomb field of a nucleus $A$. The muon pair can be produced by the Bethe-Heitler process (upper panel) and by the muon bremsstrahlung channel (lower panel). At leading order, there are four additional diagrams with identical muons exchanged. }
\label{fig:diagrams}
\end{figure}

In what follows, we will focus on the muon trident process in coherent interactions, represented by the reaction
\begin{equation}
    \mu^{\pm}_i + A \rightarrow \mu^{\pm}_f + \mu^+ + \mu^- +A,
    \label{eq:reaction}
\end{equation}
where $\mu^{\pm}_i(\mu^{\pm}_f)$ is the incoming (outgoing) muon state and $\mu^++\mu^-$ is the pair produced in the electromagnetic interaction with a nucleus $A$ with charge $Z$. At leading order, the typical Feynman diagrams associated with the muon trident scattering are represented in Fig.~\ref{fig:diagrams}. The upper diagrams represent the Bethe-Heitler contribution, where a pair of muons is produced by the two-photon fusion mechanism, with a spacelike photon being emitted by the incident muon. On the other hand, the bottom diagrams are associated to the bremsstrahlung mechanism, where a muon pair is produced by the timelike photon emitted by the incident muon. In principle, the muon production by nuclear bremsstrahlung is also possible, but the Refs.~\cite{Kelner:1998mh, Albright:1978mg, Ganapathi:1978qm} have demonstrated that this contribution is strongly suppressed. 

Over the last decades, the muon-nucleus trident cross-section has been studied theoretically within distinct formalisms and at several levels of approximation. The first analytic expression was derived by Bugaev, Kotov and Rozental (BKR) in Ref.~\cite{Bugaev1970}, extending previous results obtained for electron-positron pair production and considering only the Bethe-Heitler diagrams in Fig.~\ref{fig:diagrams}. This calculation assumed a complete screened nucleus and disregarded nuclear form factor effects, i.e., the nucleus was assumed as a point-like particle. Such approximations were discussed in Ref.~\cite{Kelner:2000va}, which have pointed out that they are expected to overestimate the total cross-section in GeV-TeV energy regime and could lead to a misinterpretation of experimental data. Additionally, Ref.~\cite{Kelner:2000va} has provided a parametrization that includes atomic and nuclear form factors. This parametrization by Kelner, Kokoulin, and Petrukhin (KKP) was implemented in the commonly used \texttt{Geant4}~\cite{GEANT4:2002zbu, Yajaman2022} package in particle physics and the \texttt{PROPOSAL}~\cite{Koehne:2013gpa, Dunsch:2018nsc} package in astroparticle physics to simulate the propagation of muons through the matter. Here, we will perform a full calculation of the cross-section, taking into account of all diagrams and masses, the interference contribution, Coulomb corrections and a precise treatment of the nuclear form factor, with the results being implemented in a Monte Carlo generator. In what follows, we will first review the approximated approaches usually present in the literature in subsection~\ref{subsec:simplified} before describing our calculation in subsection~\ref{subsec:full}.

\subsection{Simplified approaches}
\label{subsec:simplified}

The lepton-pair production at high energies is usually estimated using the Equivalent Photon Approximation (EPA) \cite{vonWeizsacker:1934nji, Williams:1934ad, Budnev:1975poe, Bertulani:1987tz}. Within this formalism, we can factorize the muon-nucleus cross-section as the product between the photon flux associated with the muon ($\gamma_\mu$) and the nucleus ($\gamma_A$), and the photon-photon fusion cross-section ($\sigma_{\gamma \gamma \rightarrow \mu^+\mu^-}$) for muon pair production as follows
\begin{equation}
\begin{aligned}
    \sigma_{\mu^{\pm} A \rightarrow \mu^{\pm} \mu^+\mu^- A}  &=  \!
    \int \mathrm{d}x_\mu \mathrm{d}x_A\mathrm{d}k_{\bot, \mu}^{2}\mathrm{d}k^{2}_{\bot, A} \\ 
    & \!\!\!\!\!\!\!\!\!\!\!\! \times  
    \gamma_\mu (x_\mu, k_\mu^2)\cdot\gamma_A (x_A,k_A^2)\cdot 
    \sigma_{\gamma \gamma \rightarrow \mu^+\mu^-} , 
\end{aligned}
\label{eq:cs_EPA_gamma-gamma}
\end{equation}
where $x_\mu(x_A)$ is the energy fraction of the photon with transverse momentum $k_{\bot, \mu}(k_{\bot, A})$ exchanged by the muon (nucleus). The photon flux associated with the muon is given by
\begin{equation}
\gamma_{\mu}(x,k_\bot^{2}) = \frac{\alpha}{2\pi}\frac{1 + (1 - x)^2}{x} \frac{k_\bot^{2}}{(k_\bot^{2}+x^2m^2)^2}, 
\end{equation}
while the photon flux from the nucleus is further modified by a factor $Z^2 \, |F(k_A^2)|^2$ with $F(k_A^2)$ being the nuclear form factor.
The cross-section $\sigma_{\gamma \gamma \rightarrow \mu^+\mu^-}$ for the lepton pair production through two photon fusion is given by the Breit-Wheeler equation~\cite{Breit:1934zz}
\begin{equation}
\label{eqn: VBS_subprocess}
\!\sigma_{\gamma \gamma \rightarrow \mu^+\mu^-} \! = \frac{2\pi\alpha^2}{s}\Bigg[\!(3 - \beta_{\mu}^4)\ln{\frac{1 \!+\! \beta_{\mu}}{1 \!-\! \beta_{\mu}}} - 2\beta_{\mu}\big(2 - \beta_{\mu}^2 \big)\!\Bigg] , 
\end{equation}
with $\beta_\mu^2= 1-4m^2_\mu/s$ and $s$ being the square of the center of mass energy of the $\gamma\gamma$ collision system. The results derived assuming this approximation will be denoted EPA($\gamma \gamma$) hereafter.

Another possibility to estimate the trident cross-section is to consider the nucleus rest frame and apply the EPA to estimate the photon flux associated with the incident muon. In this case, the cross-section will be expressed in terms of $\gamma_\mu$ and the photonuclear cross-section for the $2 \rightarrow 3$ process, ${\gamma A \rightarrow \mu^+\mu^- A}$, as follows
\begin{equation}
    \sigma_{\mu A} = 
    \int \mathrm{d}x_\mu \mathrm{d}k_\mu^{2} \,
    \gamma_\mu (x_\mu, k_\mu^2) \,
    \sigma_{\gamma A \rightarrow \mu^+\mu^- A} \, ,
    \label{eq:cs_EPA_gamma-nucleus}
\end{equation}
where we use the cross-section derived in Ref.~\cite{Jentschura:2009mb} 
\begin{equation}
    \!\!\!\sigma_{\gamma A \rightarrow \mu^+\mu^- A} \!= \!
    \frac{28}{9}\! \frac{Z^2 \alpha^3}{m_\mu^2} \!\left[ \text{ln} \frac{2 E_\gamma \Lambda}{m_\mu^2} \!-\! \frac{57}{14} \!-\! \frac{12}{35}\left(\!\frac{\Lambda}{2m_\mu}\!\right)^{\!\!2} \!\right] \!,
\end{equation}
with $E_\gamma$ being the photon energy. 
This expression assumes a monopole form factor with a root mean square radius of the tungsten nuclei given by $\Lambda=80$~MeV. The results obtained using this approximation will be denoted by EPA($\gamma A$).

The use of EPA has some limitations that must be carefully analyzed. This approach is valid for ultra-relativistic particles emitting photons, in the limit of low-$Q^2$ for the virtuality of the  exchanged  photon,  as well as when the  projectile energy carried by the emitted photon is small. In the nucleus rest frame and projectile muons in GeV to TeV energy regime, not all these conditions are necessarily satisfied, and the resulting predictions for the total cross-section are expected to overestimate the full calculation.

\begin{figure}[t]
\centering
\includegraphics[width=0.3\textwidth]{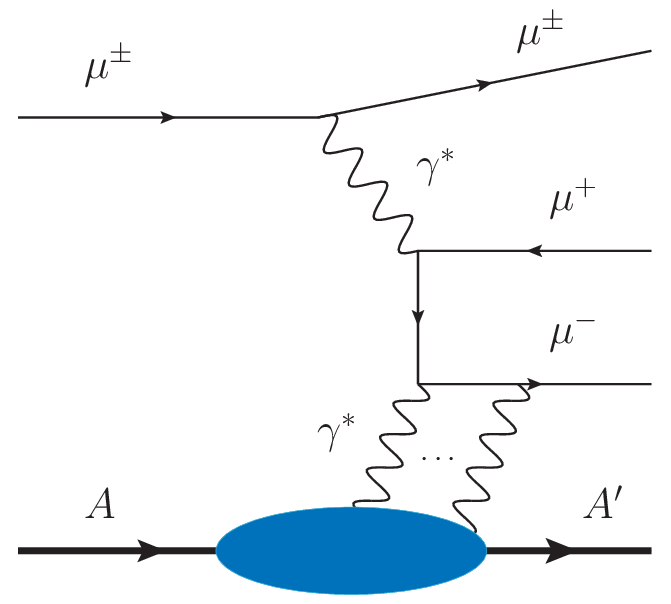} 
\caption{Coulomb corrections for the muon trident process associated with multiple photon exchanges.}
\label{fig:diagrams_CC}
\end{figure}

The Feynman diagrams presented in Fig.~\ref{fig:diagrams}, characterized by one-photon exchange, correspond to the trident amplitude in the Born approximation. However, for a heavy nuclear target, Coulomb corrections associated with multiple photon exchanges, as represented in Fig.~\ref{fig:diagrams_CC}, can become important~\cite{Davies:1954zz, Baltz:2009fs, Jentschura:2009mb, Li:2023yjt}. We will estimate the impact of the Coulomb corrections using the formalism developed in Ref.~\cite{Jentschura:2009mb}, which will be implemented in our calculations. In particular, the Coulomb correction can also be evaluated within the EPA framework, and its contribution to the photonuclear cross-section is given by~\cite{Jentschura:2009mb} 
\begin{equation}
    \sigma_{\gamma A}^\mathrm{CC} = -\frac{4332}{4375}
    \frac{Z^2 \alpha^3}{m_\mu^2} (Z\alpha)^{2} \left(\!\frac{\Lambda}{2m_\mu}\!\right)^{\!\!2}\, .
    \label{eq:cs_EPA_gamma-nucleus-CC}
\end{equation}
Therefore, the Coulomb corrections are predicted to decrease the trident cross-section.

\subsection{Full calculation}
\label{subsec:full}

In our analysis, the matrix elements associated with the Feynman diagrams for the muon trident process were estimated using \texttt{FeynCalc}~\cite{Mertig:1990an, Shtabovenko:2016sxi, Shtabovenko:2020gxv} without approximations, and the cross-section was estimated considering all masses and the interference contribution. For coherent scattering, we will assume the nucleus target is a spin-0 particle. The differential cross-section for the muon trident process in terms of the squared matrix element is then given by
\begin{equation}
\mathrm{d}\sigma = \frac{ |\mathcal{M}|^{2}}{4 M_A E_i} d \Pi_\text{LIPS},
\label{eq:differential_CS}
\end{equation}
where $\Pi_\text{LIPS}$ is the Lorentz-invariant phase-space measure of the four-particle final state. The corresponding results were implemented in the Monte Carlo (MC) generator presented in Ref.~\cite{Francener:2025wzh}, which has generalized the Monte Carlo developed in Ref.~\cite{Altmannshofer:2019zhy}. This generator performs the phase-space integration over seven kinematic variables and generates events using a rejection-sampling algorithm. 

The MC generator includes tungsten and lead targets, corresponding respectively to the target composition of the FASER and NA64$\mu$ experiments. Moreover, a precise description of these nuclei was implemented through the nuclear form factor parametrized by a Fourier transform of the Woods-Saxon nuclear charge distribution~\cite{Woods:1954zz} with parameters obtained from~\cite{DeVries:1987atn}. In addition to the coherent muon-nucleus scattering, the MC generator can also calculate incoherent scatterings with proton and neutron targets inside a heavy nucleus. In such case, the nucleus is modeled as a Fermi gas of protons and neutrons, including a Pauli blocking factor for the outgoing nucleons \cite{Bell:1963ogq}.

\begin{figure*}[t]
	\centering
	\begin{tabular}{c}
\includegraphics[width=1\textwidth]{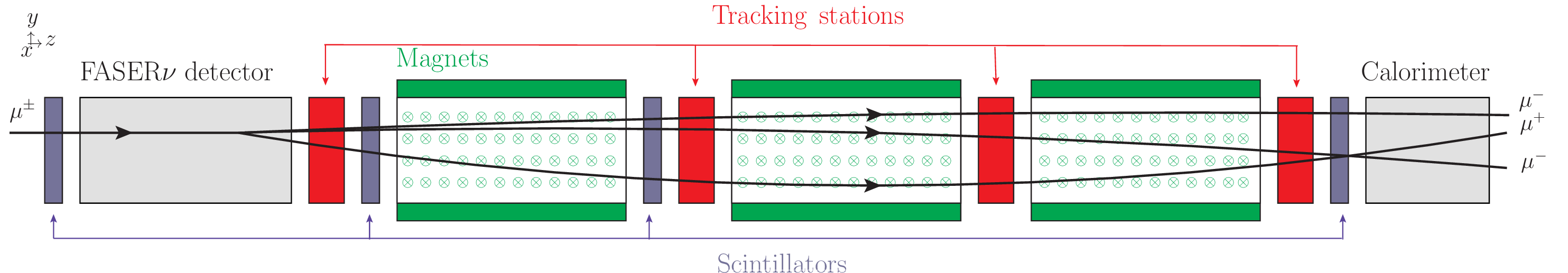} 
	\end{tabular}
\caption{Schematic side view of FASER experiment with a muon undergoing a trident scattering in the tungsten plates and three muons being identified by the FASER spectrometer.}
\label{fig:experiment}
\end{figure*}

As previously mentioned, nuclear form factors play an important role in the muon trident process, and assuming the nucleus as a point-like particle can overestimate the cross-section by up to one order of magnitude depending on the energy regime. To estimate the dependence of our predictions on this ingredient, different models to describe the nucleus target were implemented in the Monte Carlo generator used in our analysis. In particular, the nucleus as a point-like Coulomb center has been implemented, as well as a more realistic description based on the numerical Fourier transform of the Woods-Saxon nuclear charge distribution. In addition, two other models commonly found in the literature were also implemented: the monopole form factor, derived from a Yukawa nuclear charge distribution~\cite{Yukawa:1935xg}, and the often referred as realistic form factor, obtained considering the nucleus as a convolution of a hard sphere with the Yukawa nuclear charge distribution~\cite{Klein:1999qj}. These two phenomenological form factors have analytical expressions, which are presented in Appendix~\ref{appendix:FormFactor}. In addition to the nuclear form factor, we will also consider the atomic form factor, associated with a nucleus screened by electrons. The parametrization for heavy nuclei presented in \cite{Schiff:1951zza, Tsai:1973py} has been implemented. We have verified that, differently from the nuclear form factors, the impact of the inclusion of the atomic form factor on our predictions is small.

\section{Experimental Setup}
\label{sec:experimental}

In what follows, we will describe the FASER and NA64$\mu$ detectors, the muon flux that reaches it, and the settings of our simulations for the muon trident scattering, including acceptance cuts and the potential backgrounds for the measurement of the muon trident process.

\subsection{FASER Detector}

A schematic picture of the FASER detector~\cite{FASER:2018bac} is presented in Fig.~\ref{fig:experiment}. Located at the front is FASER$\nu$, a dedicated neutrino detector consisting of 730 alternating layers of 1.1~mm thick tungsten plates and emulsion films~\cite{FASER:2019dxq, FASER:2020gpr}. This is followed by FASER's spectrometer~\cite{FASER:2021ljd} consisting of four tracking stations and three air-filled magnets with an inner aperture of 10~cm and a field strength of 0.55~T. In addition, the detector contains several scintillators, including before and after FASER$\nu$. Placed at the downstream end is an electromagnetic calorimeter. 

The tungsten of FASER$\nu$ will constitute the target for muon trident interactions. As discussed in Ref.~\cite{Francener:2025wzh}, it is possible to detect and study these interactions solely with the emulsion detector by measuring the muon momenta via multiple-Coulomb scattering~\cite{FASER:2025qaf, FASER:2026hzm}. The full analysis of the emulsion data is, however, a labor and computation-intensive process, and the complete dataset will therefore only become available for physics analyses gradually. In this study, we will follow a different strategy and propose the detection of muon tridents using only the electronic detector components. A muon trident candidate event is required to consist of three muon tracks passing through the spectrometer. In addition, we require a single MIP-like (Minimum Ionizing Particle like) charge deposit in the scintillator before tungsten target and three MIP-like charge deposit in the scintillators after FASER$\nu$. 

We estimate the event rate of muon tridents occurring in FASER$\nu$ during LHC Run~3 with an integrated luminosity of $\mathcal{L}_\mathrm{pp} = 350~\mathrm{fb}^{-1}$. The muon flux reaching the FASER$\nu$ detector used in our simulation was obtained from FLUKA simulations~\cite{Battistoni:2015epi, Sabate-Gilarte:2023aeg} and made available in LHAPDF format~\cite{Buckley:2014ana} in Ref.~\cite{Francener:2025pnr}. We note that flux assumes the configuration of the LHC at the beginning of LHC Run~3. Due to changes of the beam crossing angle direction and beam optics, the muon flux observed at FASER increased in later years. In this sense, our estimate of the event rate should be considered conservative. We emphasize, however, that the uncertainty in the muon flux can be disregarded, given that muon flux can be measured in situ with FASER spectrometer and the muon flux simulations validated independently of cross-sections.

In this work, we are considering the muons reaching FASER$\nu$ within a circular area of radius 10~cm, corresponding to the geometrical overlap between FASER$\nu$ and FASER spectrometer. Under the assumptions described above, a flux of $3\times 10^9$ muons and anti-muon with an energy of more than 10~GeV is expected to pass through the considered target volume. We simulate muon trident interactions distributed uniformly throughout the target volume and simulate propagation of muons through the following spectrometer. To permit a clean reconstruction and momentum measurement, we require the muons to stay within the spectrometers acceptance for at least the first three tracking stations. This in particular reduces the efficiency to detect for low-energy muons, which are more easily bend out of the spectrometer's acceptance. We note that it may, in principle, be possible to reconstruct low-energy muons with just the first two tracking stations at the cost of a poorer momentum resolution.  

We can estimate the momentum resolution $\sigma_p$ of a particle with momentum $p$ using the Gluckstern formula
\begin{equation}
    \frac{\sigma_p}{p} = \frac{\sigma_x \cdot p}{e B L^2} \left( \frac{720}{N+4} \right)^{1/2},
\end{equation}
where $\sigma_x \sim 20\,\mu$m is the spatial resolution of the tracking layer in the precision direction, $L=2.5$~m is the length of the track in the magnetic field $B=0.55$~T, and $N=9$ is the number of tracking layers in the first three stations. Using this assumptions we obtain $\sigma_p/p \approx 20\% \times (p/\text{TeV})$. High energy muons typically pass through all four tracking stations. In this case, the momentum resolution further improves by about a factor of two.  While the actual momentum resolution is more complex, owing for example to alignment uncertainties, this simple estimate illustrates that the spectrometer is capable of measuring the final-state muon momenta with sufficient precision to study muon tridents over the full energy range.

\subsection{NA64$\mu$ Detector}
In addition to the broadband muon beam emerging from $pp$ collisions at the LHC, the muon trident process can also be studied using the using the M2 beamline at CERN Super Proton Synchrotron,  which provides a high intensity muon beam, although at lower energy. One of the experiments is the NA64$\mu$ \cite{Gertsenberger:2023ree}, whose primary purpose is the exploration of dark sector physics via missing transverse momenta in muon scattering (see, e.g., Ref.~\cite{Chen:2018vkr}). Such a fixed-target experiment uses  a 160 GeV muon beam with $3.7\%$ incoming beam energy resolution~\cite{NA64:2018iqr} and is equipped with a spectrometer before/after target. In our analysis, we will also estimate the muon trident cross-section at the NA64$\mu$ experiment \cite{NA64:2018iqr}, assuming that it is operating with a 160~GeV muon beam and a total of $10^{13}$ muons on target, which is expected for its phase II \cite{NA64:2018iqr}. For the acceptance cuts, we will assume the experimental design described in Ref.~\cite{NA64:2024nwj}, and a trident candidate is assumed to have three muons in the final state crossing both hadronic calorimeters (HCAL$_{1,2}$), and a minimum of 10~GeV from each final muon. As we will show below, the statistics will not be a limitation for the analysis of the trident process at the NA64$\mu$ experiment.

\subsection{Backgrounds}
\label{subsec:backgrounds}

In this subsection, we summarize the main backgrounds associated with muon trident. Since our analysis does not rely on the reconstruction of the interaction vertex within the tungsten plates using the emulsion detector, the dominant backgrounds are regard to processes producing three muons exiting FASER$\nu$ and entering the spectrometer. In what follows, we summarize the most important process with three muons in the final state.

\begin{description}
    \item [Photon initiated Showers] This process is characterized by muon bremsstrahlung while crossing tungsten plates, and subsequent photon conversion in a pair of muons through the reaction
    \begin{equation}
    \mu^{\pm}_i \!+\! W \rightarrow \mu^{\pm}_f  \gamma^* \!+\! W  \rightarrow \mu^{\pm}_f  \mu^+  \mu^- \!+\! W \!+\! X\, .
    \end{equation}
    To estimate the event rate associated with this process, we simulate muon bremsstrahlung and its electromagnetic shower with the generator developed by the authors in Ref.~\cite{Kling:2025lnt}. In addition to the muon pair coming from the photon radiated by the muon, the simulation takes into account the contribution from the electromagnetic shower evolution, where a pair of muons can come from photons radiated by electrons and positrons. 

    \item[Electron Initiated Shower] 
    In this process, an electron-positron pair is produced through Bethe-Heitler or bremsstrahlung channels from Fig.~\ref{fig:diagrams}, and the muon pair comes from the photon conversion in the evolution of the electromagnetic shower produced, in a reaction given by
    \begin{equation}
    \mu^{\pm}_i \!+\! W \rightarrow 
    \mu^{\pm}_f  e^+ \!+ e^-  W \rightarrow 
    \mu^{\pm}_f  \mu^+  \mu^- \!+\! W \!+\! X \, .
    \end{equation}
    This process was estimated with the electron-positron spectra obtained from Ref.~\cite{Francener:2025wzh}, and the electromagnetic shower was again simulated with the generator proposed in Ref.~\cite{Kling:2025lnt}.

    \item [Muon DIS] This background is constituted from charged hadrons produced in muon deep inelastic scattering (DIS) misidentified as muons when measured by the electronic detector through the reaction
    \begin{equation}
    \mu^{\pm}_i + W \rightarrow \mu^{\pm}_f +  h_1 + h_2 + X \, .
    \end{equation}
    In order to estimate this contribution, we simulated muon DIS interactions at FASER$\nu$ using the \texttt{POWHEG-BOX-RES} generator \cite{Banfi:2023mhz, vanBeekveld:2024ziz, FerrarioRavasio:2024kem} and the hadronization with \texttt{Pythia~8}~\cite{Bierlich:2022pfr}, as done in Ref.~\cite{Francener:2025pnr}. After hadronization process, we are propagating hadrons through the tungsten using \texttt{Pythia~8} to describe hadron-hadron interactions. For simplicity, we are not propagating the associated electromagnetic shower initiated by neutral pions produced in DIS. We are considering that all kind of charged hadrons can be identified as muons, and we are taking into account only events with two charged hadrons with momentum larger than 3~GeV reaching the scintillator at the end of FASER$\nu$.
    
    \item [Muon DIS with Rare Meson Decays] Muons can be produced in rare meson decays, such as the decay $\eta \to \mu\mu\gamma$ where the $\eta$ was produced in muon DIS. The corresponding reaction chain is
    \begin{equation}
    \mu^{\pm}_i + W \rightarrow \mu^{\pm}_f  \eta + X \rightarrow \mu^{\pm}_f  \mu^+  \mu^- + X \, .
    \end{equation}
    In order to estimate this contribution, we again simulated muon DIS interactions at FASER$\nu$ using the \texttt{POWHEG-BOX-RES} generator and the hadronization with \texttt{Pythia~8}. With this simulation, we have found that approximately 40\% of muon DIS processes at FASER$\nu$ have at least one $\eta$ meson in the final state. However, the contribution to the trident background is strongly suppressed by the small branching ratio $\mathrm{Br}_{\eta \rightarrow \mu\mu \gamma} = (3.1\pm 0.4)\times 10^{-4}$ ~\cite{ParticleDataGroup:2018ovx}. Overall, we find that this background is suppressed by about a factor hundred compared to the previously discussed electromagnetic showers. We neglect contributions from the additional light meson decays into muons, such as $\omega, \rho, \phi \to \mu \mu$, which  are even expected to be subdominant due to their smaller branching fractions into muons and/or lower production rates. 

    \item [Muon DIS with Charm Decay] Muons can also be produced in charm hadron decays, where a charm hadron pair was produced in muon DIS. This contribution was also estimated with \texttt{POWHEG-BOX-RES} to simulate the muon DIS with tungsten and \texttt{Pythia~8} to simulate the hadronization process. This process can be a background to muon trident when both charm hadrons decay into muons:
    \begin{equation}
    \mu^{\pm}_i \!+\! W \rightarrow \mu^{\pm}_f  c  \bar{c} \!+\! X \rightarrow \mu^{\pm}_f \mu^+ \mu^- \!+\! X \, .
    \end{equation}
    This process is suppressed by the square of the branching ratio $\mathrm{Br}_{c \rightarrow \mu + X} = 0.096\pm 0.04$ \cite{ParticleDataGroup:2018ovx} as well as by the charm parton distribution function. Overall, we find that this background is suppressed by  a factor of about one hundred compared to the previously discussed electromagnetic showers.

    \item [Multiple muons] As the muon beam intensity is large, multiple muons can enter the FASER$\nu$ experiment. Such a background can be rejected using the front veto scintillators by requiring a 1MIP like charge deposit. In principle, muons could also enter undetected from the side and scatter. This effect was studies in FASER's muon neutrino analysis~\cite{faser_collaboration_2026_eg0mj-1ep53}, which found that this contribution is negligible for $p>50$~GeV.
\end{description}

In the next section, we will present differential distributions that may help suppress background contributions by selecting specific kinematic regions.

\section{Results}
\label{sec:results}

One of the main goals of this study is to investigate the potential of the FASER detector to measure the muon trident scattering with TeV energy muons from the LHC for the first time. In what follows, we will present our results for the projected sensitivity for future measurements comparing different predictions for this process that have been used in the literature. We will also discuss differential distributions and how they can help us to test models as well as to separate the muon trident signal from its associated background. However, before to present these results, it is important to analyze the dependence of the predictions on the underlying assumptions of the theoretical calculation of the total cross-section.

\subsection{Comparison between theoretical predictions}

As discussed in the previous section, over the last decades the muon trident cross-section have been derived assuming distinct approximations and assumptions for the nuclear form factor. In particular, the Bugaev, Kotov and Rozental \cite{Bugaev1970} have considered a point-like nuclear form factor in its calculations and obtained an analytical result. On the other hand, Kelner, Kokoulin, and Petrukin \cite{Kelner:2000va} have considered a nucleus with the Woods-Saxon charge distribution and have also derived an analytical formula that has been implemented in the \texttt{Geant4} and \texttt{PROPOSAL} software packages.  As already demonstrated in Fig.~\ref{fig:projected_sensitivity}, the full predictions derived in this work agree at low energies with those derived using the \texttt{Geant4} and \texttt{PROPOSAL} event generators. However, we predict larger cross-sections in the kinematical region that will be probed by the FASER and NA64$\mu$ experiments. At TeV energies, \texttt{Geant4} (\texttt{PROPOSAL}) predicts $\sim 65\%$ ($\sim 25\%$) of our prediction.  As a consequence, future experimental data will, in principle, be able to validate (or not) the theoretical description of the muon trident process implemented in the available Monte Carlo generators.

\begin{figure}[t]
	\centering
	\begin{tabular}{c}
\includegraphics[width=0.47\textwidth]{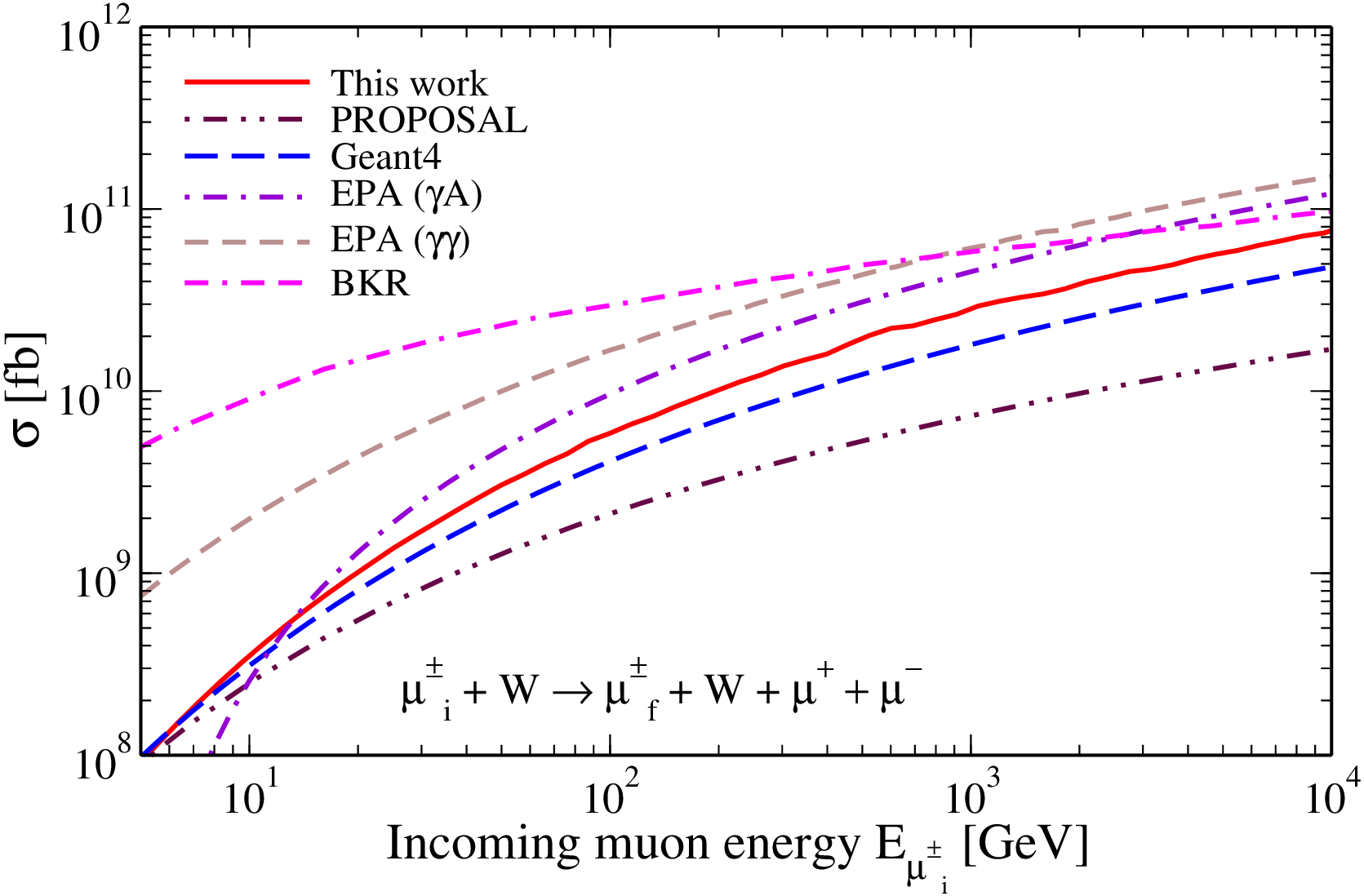} 
	\end{tabular}
\caption{ Muon trident cross-section as a function of incident muon energy for a tungsten target derived considering different approximations and distinct approaches. }
\label{fig:cross_section_muon_trident_Models}
\end{figure}

In Fig.~\ref{fig:cross_section_muon_trident_Models} we compare the BKR and KKP predictions with those derived assuming  the Woods-Saxon nuclear charge distribution,  and the Equivalent Photon Approximation to describe the muon pair production by $\gamma \gamma$ and $\gamma A$ interactions,  discussed in the previous section. The KKP prediction have been calculated using the \texttt{PROPOSAL} event generator for a tungsten target.  In agreement with previous results, the BKR prediction overestimates the KKP one, especially for low energies.  On the other hand, the EPA predictions overestimate our predictions by a factor $\approx 2$ in the TeV regime, which demonstrated that such an approach is not a good approximation to estimate the total cross-section for the muon trident process at these energies.
In what follows, we will focus on the results derived using our Monte Carlo generator, described in subsection \ref{subsec:full}.

\subsection{Total event rates}

\begin{table}[b]
\centering
\renewcommand{\arraystretch}{1.5}
\begin{tabularx}{\linewidth}{XXXl}
\hline \hline
\multicolumn{4}{c}{\bf Events for muon trident scattering at FASER}\\
\hline
                       & {\bf Incoherent}    & {\bf Coherent}        & {\bf Total} \\
\hline	
Without cuts           & $ 1.9 \times 10^3 $ & $ 226.9 \times 10^3 $ & $ 228.8 \times 10^3 $ \\
With cuts              & $ 391.5 $           & $ 85.2 \times 10^3 $  & $ 85.6 \times 10^3 $ \\
\hline \hline
\end{tabularx}
\caption{ Predictions for the number of muon trident scattering at the FASER assuming an integrated luminosity of 350 fb$^{-1}$ during Run~3 of the LHC. We are considering events in the overlap between FASER$\nu$ and FASER electronic detector.
}
\label{table:Nevents_total_Faser}
\end{table}

\begin{table}[t]
\centering
\renewcommand{\arraystretch}{1.5}
\begin{tabularx}{\linewidth}{XXXl}
\hline
\hline
\multicolumn{4}{c}{\bf Events for muon trident scattering at NA64$\mu$}\\
\hline
                       & {\bf Incoherent}    & {\bf Coherent}        & {\bf Total} \\
\hline	

Without cuts           & $ 6.15 \times 10^5 $ & $ 8.19 \times 10^7 $ & $ 8.25 \times 10^7 $ \\

With cuts              & $ 2.21 \times 10^4 $ & $ 8.77 \times 10^6 $  & $ 8.79 \times 10^6 $ \\
\hline \hline
\end{tabularx}
\caption{ Predictions for the number of muon trident scattering at the NA64$\mu$ assuming $10^{13}$ muons on target (MOT). }
\label{table:Nevents_total_NA64}
\end{table}

In Table \ref{table:Nevents_total_Faser} we present our predictions for the muon trident process at FASER.  We are considering in our analysis only a circular area of FASER$\nu$ with $r<0.1\;\mathrm{m}$ for muon interactions. This region corresponds to the geometrical overlap between FASER$\nu$ and FASER spectrometer that should be used for the identification of outgoing particles. The muon-tungsten cross-section is estimated considering the Woods-Saxon nuclear charge distribution to model the tungsten form factor in the coherent scattering. We are also presenting predictions for incoherent scatterings of individual nucleons inside tungsten nucleus. We present predictions for both the total event rates and the number of events surviving the FASER tracking acceptance cuts. After acceptance cuts, $\simeq20\%$ of muon trident events at FASER have three muon tracks inside FASER spectrometer, corresponding to more than $8.5\times10^4$ events. Incoherent scatterings represent about 0.8\% of total events before cuts, and after cuts its contribution decreases to 0.3\%. Events with incoherent scatterings are more strongly affected by acceptance cuts because they typically have muons with larger transverse momenta than those produced in coherent process, thus they leave the tracking system more easily. In contrast, the results for the NA64$\mu$ experiment, presented in Table \ref{table:Nevents_total_NA64}, indicate that the expected number of events under the above assumptions are $8.2\times 10^7$ and $8.8\times 10^6$ before and after the acceptance cuts adopted for coherent scatterings. The large number of events at NA64$\mu$ suggest that its uncertainties will be dominated by systematics.

In Table \ref{table:Nevents_background_Faser}, we present the muon trident signal together with the contributions associated with each background channel discussed above for events with three muons in the final state. Our results indicate that muon bremsstrahlung, $e^+e^-$ shower, and muon DIS with charged hadrons identified as muons are the main background sources for muon trident scattering, together corresponding to approximately 14\% of the signal. Muon DIS with $\eta$ or charm quarks in the final state are strongly suppressed, being less than 0.3\% of the signal. Similar percentages for the backgrounds have been obtained for the NA64$\mu$ experiment. The backgrounds originating from muon DIS can also be more suppressed if the spectrometer can reconstruct additional charged particles produced in the interaction. Likewise, background events containing an associated EM shower can be further suppressed. 

\begin{table}[b]
\centering
\renewcommand{\arraystretch}{1.5}
\begin{tabularx}{\linewidth}{Xr}
\hline
\hline
\multicolumn{2}{c}{\bf Estimative of backgrounds for muon trident }\\
\hline
 {\bf Process}     & {\bf Number of events}  \\
\hline	 
Trident signal & $ 85.2 \times 10^3 $  \\
Muon bremsstrahlung & $ 3.4 \times 10^3 $   \\
$e^+ + e^-$ pair production & $ 7.6 \times 10^3 $   \\
Muon DIS & $ 1.3 \times 10^3 $   \\
DIS with $\eta$ & $ < 27.2 $   \\
DIS with charm & $ < 144.3 $   \\
\hline
\hline
\end{tabularx}
\caption{ Main backgrounds at FASER for muon trident process that produce three muons in the final state.  }

\label{table:Nevents_background_Faser}
\end{table}

\begin{figure*}[t]
	\centering
	\begin{tabular}{cc}
\includegraphics[width=0.5\textwidth]{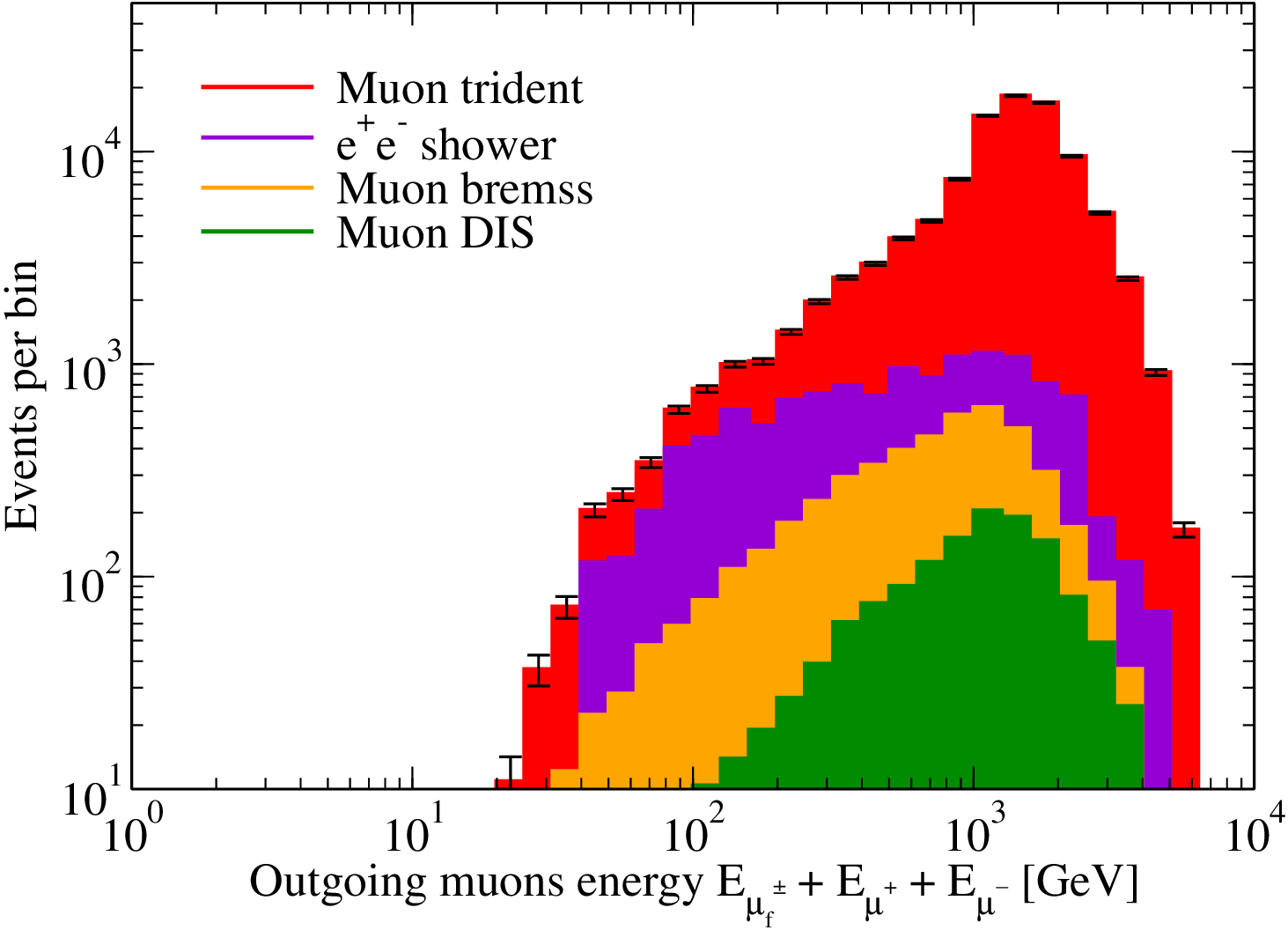} 
\includegraphics[width=0.5\textwidth]{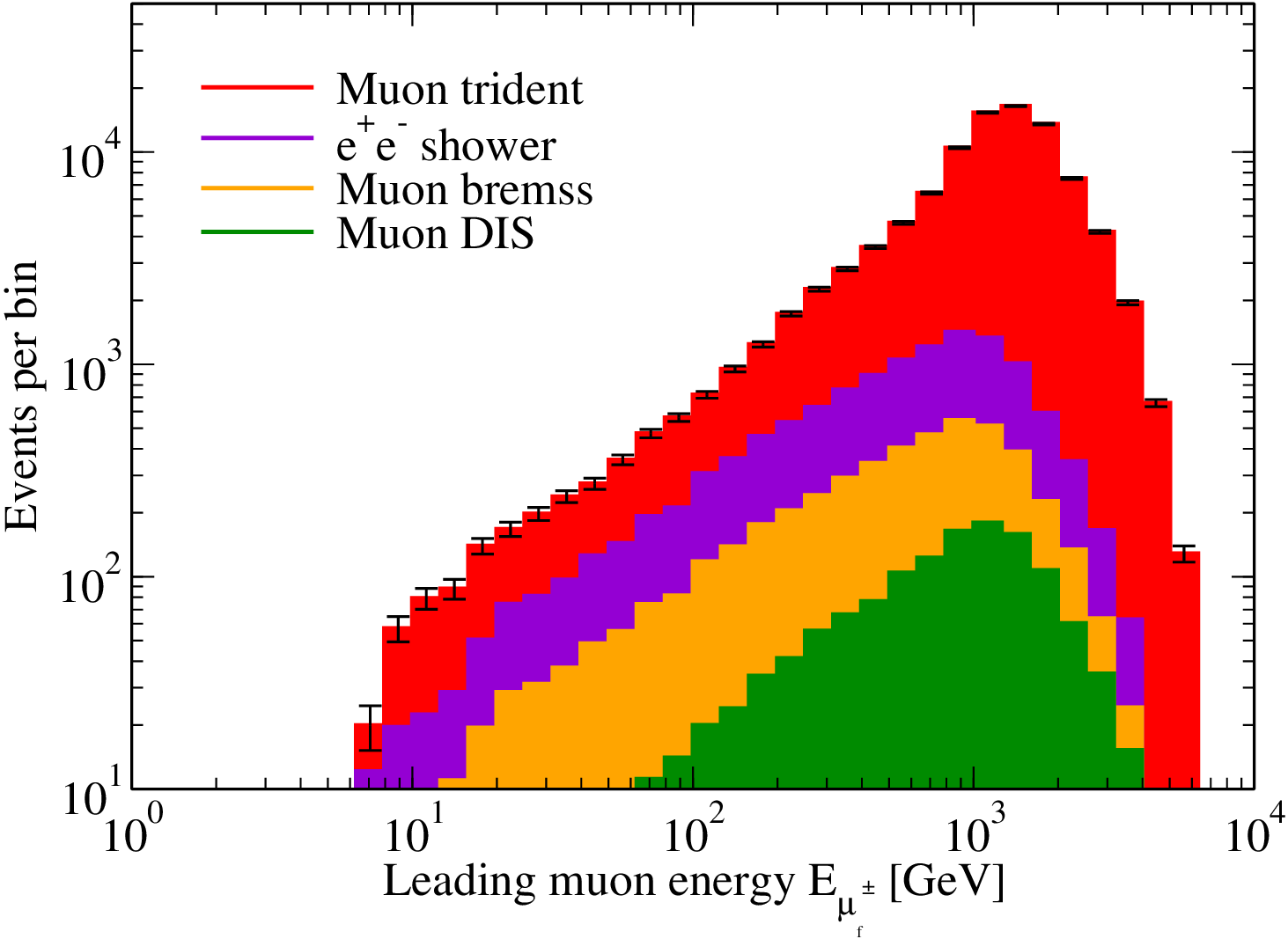} \\
\includegraphics[width=0.5\textwidth]{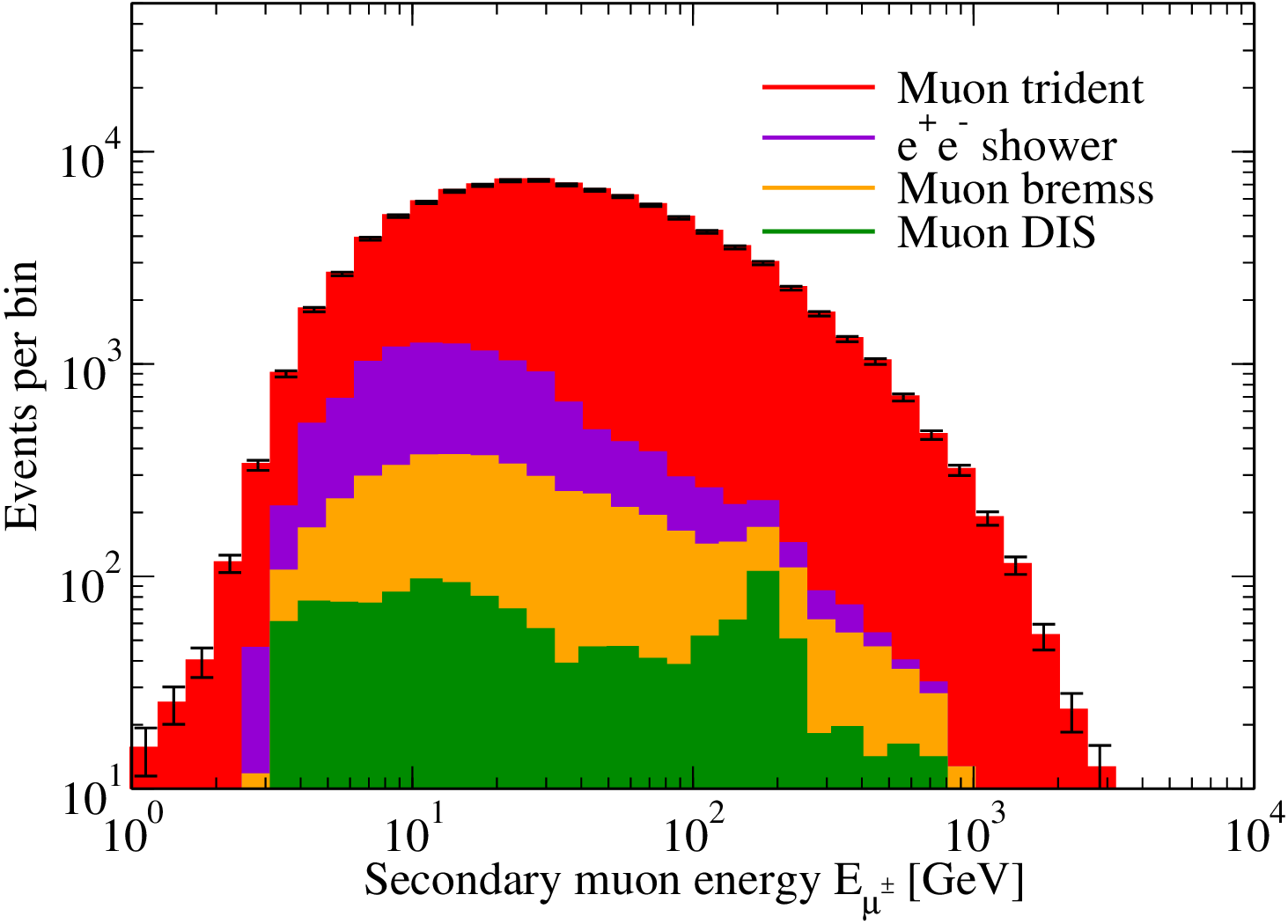} 
\includegraphics[width=0.5\textwidth]{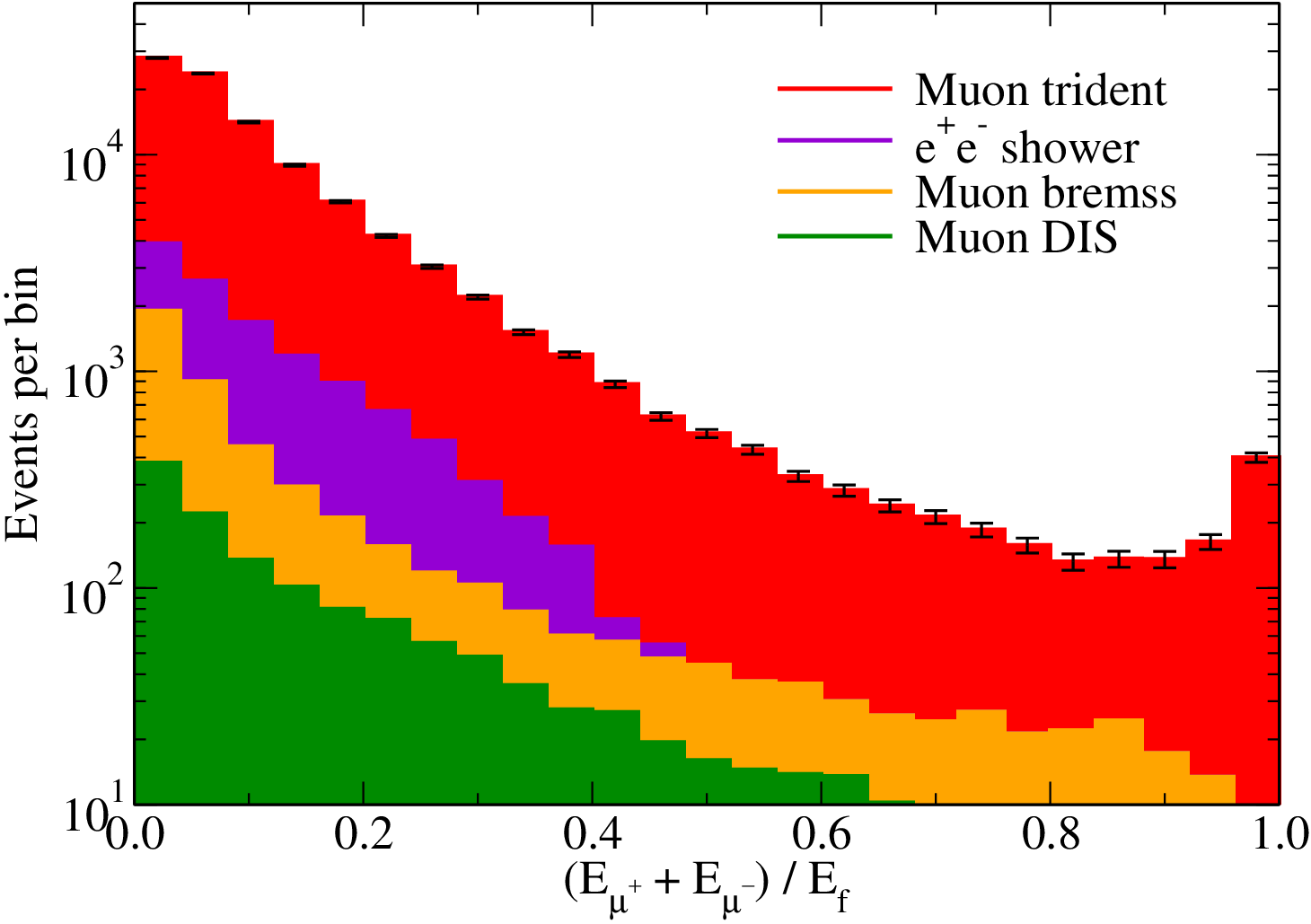} \\
\includegraphics[width=0.5\textwidth]{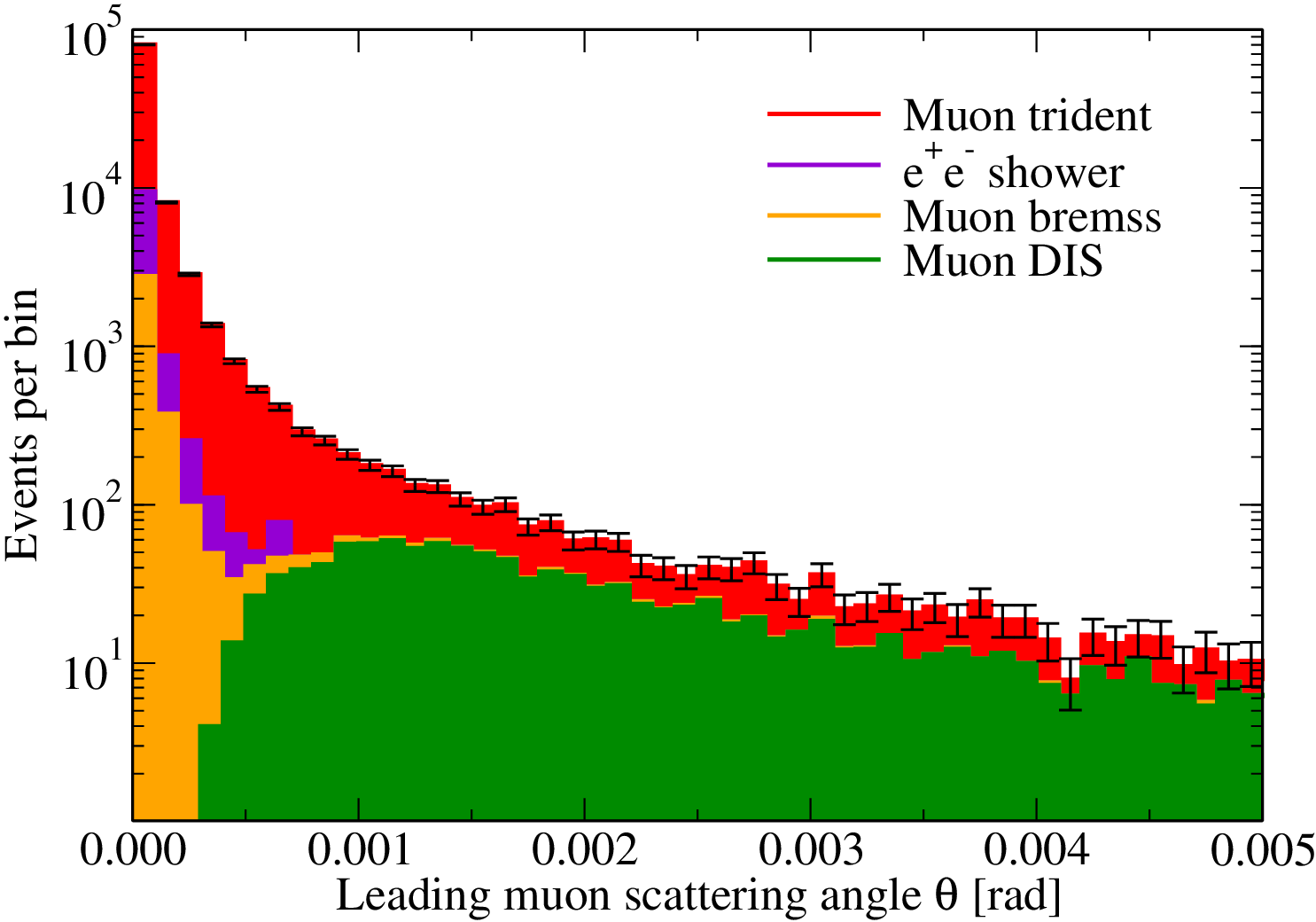} 
\includegraphics[width=0.5\textwidth]{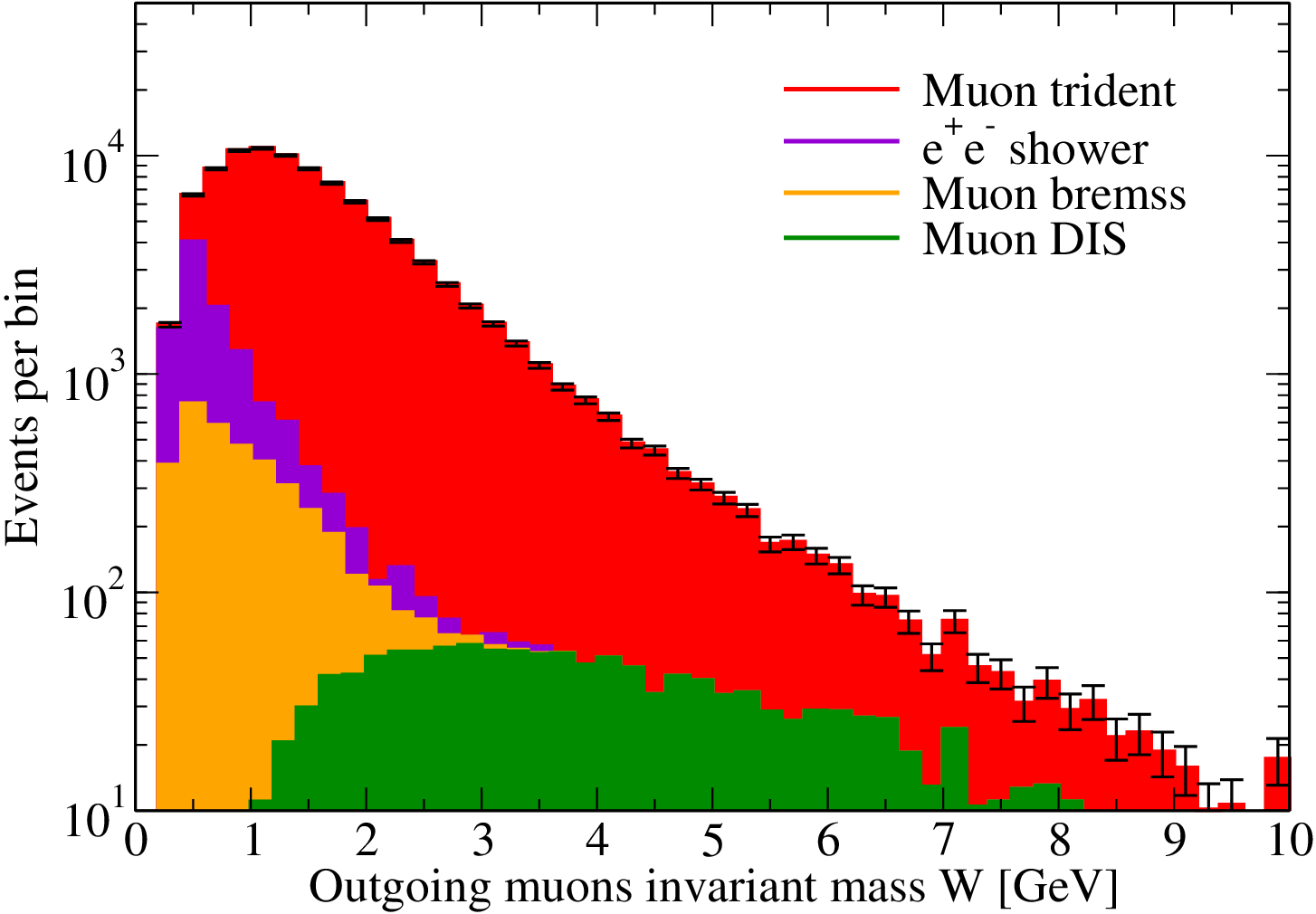} 
	\end{tabular}
\caption{ Number of events for the muon trident process as well as its main backgrounds binned in distinct configurations of energy (top and middle panels) and binned in the incident muon scattering angle (bottom left) and in the invariant mass of the three final state muons (bottom right). }
\label{fig:hist_events_bkgd}
\end{figure*}

\subsection{Differential Distributions}

In Fig.~\ref{fig:hist_events_bkgd} we present our predictions for selected differential distributions of muon trident events at the FASER detector during Run~3 of the LHC, as well as for the main backgrounds, after applying the acceptance cuts described in the previous section. Our predictions were estimated considering the Woods-Saxon nuclear charge distribution. In the top left (right) panel, we present events binned in the total energy of the three final state muons (hardest final state muon). The left panel shows signal and backgrounds with similar shapes, but a noticeably larger background/signal ratio at lower final muons energies due to the weaker energy dependence of the background cross-sections compared to the trident signal, as well as due to low energy muons produced in the evolution of the EM showers. The muon trident signal in the top left and right panels are similar, given that the incident muon energy transferred to the nucleus is small, leaving a hard muon with high energy. In contrast, the background distributions are significantly modified due to the sizable energy fraction of the incident muon transferred to the final electromagnetic and hadronic showers. 

The middle panel of Fig.~\ref{fig:hist_events_bkgd} shows events binned in the energy of a produced muon (left), which can identified as a single positive or negative muon in the final state, and the fraction of the initial muon energy transferred to the produced muons. In these plots we can observe that muons produced in the trident process typically carry only a small fraction of the incident muon energy, suggesting that the most energetic outgoing muon is typically associated with the scattered incident muon. Muons arising from $e^+e^-$ shower background are usually even less energetic than those from trident scattering.

Finally, in the bottom panels of Fig.~\ref{fig:hist_events_bkgd}, we compare the trident signal and backgrounds binned in the scattering angle of the hardest muon (left) and the invariant mass of the three final muons produced (right). The muon trident signal is characterized by small deflections of the incident muon, with more than 99\% of events having a deflection smaller than 1~mrad. This feature is also observed for the muon bremsstrahlung and $e^+e^-$ shower backgrounds; all events simulated have incident muons deflected by less than 2~mrad. On the other hand, muons from the DIS background are characterized by larger scattering angles, and this background becomes bigger than the trident signal for angles $\gtrsim 2$~mrad. The invariant mass distribution of three final muons indicates that the muon bremsstrahlung and $e^+e^-$ shower backgrounds are concentrated at smaller invariant masses than the trident signal. This behavior is associated with small deflections for produced muons, hence also small transverse momenta. The opposite behavior is observed in the muon DIS background, where the invariant mass distribution of final state particles is concentrated in larger values than the trident signal. Among the observables considered, the invariant mass and the muon scattering angles provide the strongest discrimination between signal and background.

\subsection{Dependence on the nuclear form factor}

The results presented above were obtained assuming the nucleus target described by the Woods-Saxon nuclear charge distribution. In this subsection, we compare the different nuclear form factors commonly used in phenomenological predictions. Firstly, 
in Table~\ref{table:Nevents_formfactor_faser} we present our predictions for the number of muon trident events in coherent scattering comparing different form factors. We show predictions with and without acceptance cuts. Our results indicate  that the muon trident scattering is strongly dependent on the nuclear form factor, and even the realistic one differs by approximately 5\% from Woods-Saxon predictions. Our results also indicate that events generated with simpler form factor models are more strongly affected by acceptance cuts, since these form factors predict more events with large scattering angles and transverse momenta for muons produced. These results indicate that muon trident measurements at FASER may provide sensitivity to the modeling of nuclear form factors.

\begin{table}[t!]
\centering
\renewcommand{\arraystretch}{1.5}
\begin{tabularx}{\linewidth}{XXXXl}
\hline \hline
\multicolumn{5}{c}{\bf Events for muon trident scattering at FASER}\\
\hline
                & {\bf Woods-Saxon}     & {\bf Realistic}       & {\bf Monopole}        & {\bf Point-like} \\
\hline	
Without cuts    & $ 226.9 \times 10^3 $ & $ 240.9 \times 10^3 $ & $ 258.9 \times 10^3 $ & $ 581.8 \times 10^3 $ \\
With cuts       & $ 85.2 \times 10^3 $  & $ 89.1 \times 10^3 $  & $ 91.6 \times 10^3 $  & $ 154.0 \times 10^3 $ \\
\hline
\hline
\end{tabularx}
\caption{ Muon trident events at FASER with and without acceptance cuts assuming distinct parameterizations to describe the nuclear form factor.
}
\label{table:Nevents_formfactor_faser}
\end{table}

\begin{figure}[t]
	\centering
	\begin{tabular}{c}
\includegraphics[width=0.47\textwidth]{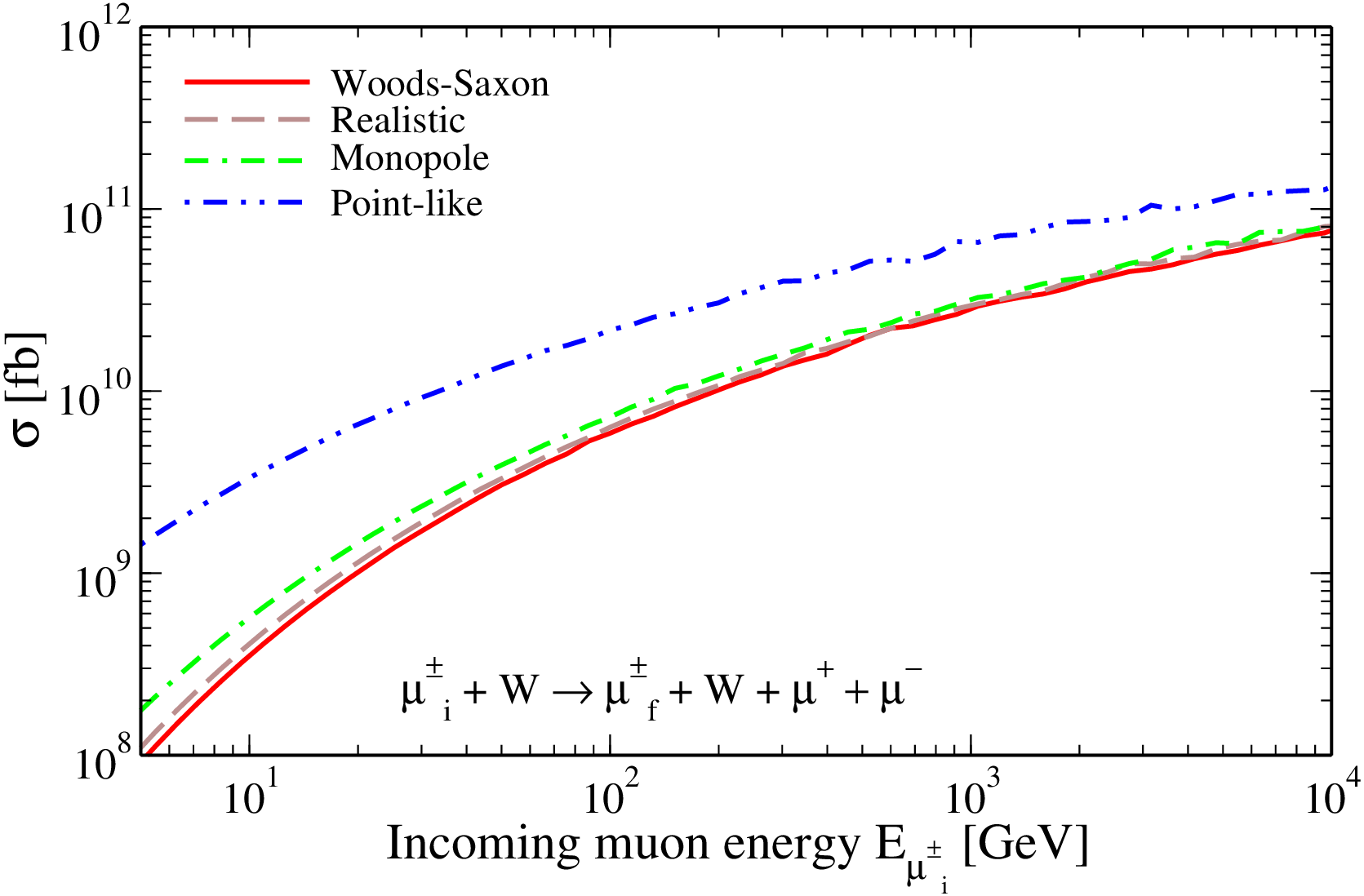}  
	\end{tabular}
\caption{ Muon trident cross-section (top) and events at FASER (bottom)  as a function of incident muon energy for distinct models of nuclear form factor. 
}
\label{fig:cs_formfactors}
\end{figure}

The dependence of the total cross-section on the nuclear form factor is analyzed in Fig.~\ref{fig:cs_formfactors}. We present the total cross-section for muon trident scattering as a function of incident muon energy, considering Woods-Saxon, realistic, monopole, and point-like form factors. All alternative parametrizations imply larger values for the  total cross-section, especially at lower energies. At the TeV scale, the realistic, monopole, and point-like form factors increase the cross-section by 2.5\%, 11.5\% and 113\%, respectively. The event distributions binned in the incident muon energy closely follow the form factor dependence observed in the total cross-section. The energy dependence on the nuclear form factor motivates the measurement of muon trident scattering through a large energy range.

\subsection{Impact of the Coulomb corrections and of the interference contribution}

As discussed in Section \ref{sec:cross_section_modeling}, the contribution of Coulomb corrections can become non-negligible for heavy nuclear targets, as those present at FASER and NA64$\mu$ experiments. In Table~\ref{table:Nevents_CC} we present events at FASER and at NA64$\mu$ experiments obtained in the first Born approximation with those including Coulomb corrections. At FASER, the expected number of events decreases by 1.4\% while at NA64$\mu$ by 6.1\% once this correction is included. Our results indicate that Coulomb corrections exhibit a nontrivial dependence on the incident muon energy. Consequently, a precise measurement of the muon trident cross-section at NA64$\mu$ ($\sim160$~GeV) and FASER (TeV scale) could provide sensitivity to these higher-order QED effects. 

\begin{table}[t]
\centering
\renewcommand{\arraystretch}{1.5}
\begin{tabularx}{\linewidth}{Xccc}
\hline\hline
\multicolumn{4}{c}{\bf Events for muon trident scattering with CC}\\
\hline
                & {\bf First Born approx.\!\!\!}     & {\bf CC}       & {\bf Total} \\
\hline	
FASER    & $ 85.2 \times 10^3 $ & $ -1.03 \times 10^3 $ & \ \ $ 84.2 \times 10^3 (-1.2\%) $  \\
NA64$\mu$& $ 8.74 \times 10^6 $ & $ -0.53 \times 10^6 $ & \ \ $ 8.21 \times 10^6 (-6.1\%) $   \\
\hline\hline
\end{tabularx}
\caption{ Impact of Coulomb corrections in muon trident events at the FASER and at the NA64$\mu$ experiments after acceptance cuts.
}
\label{table:Nevents_CC}
\end{table}

\begin{figure}[t]
	\centering
	\begin{tabular}{c}
\includegraphics[width=0.48\textwidth]{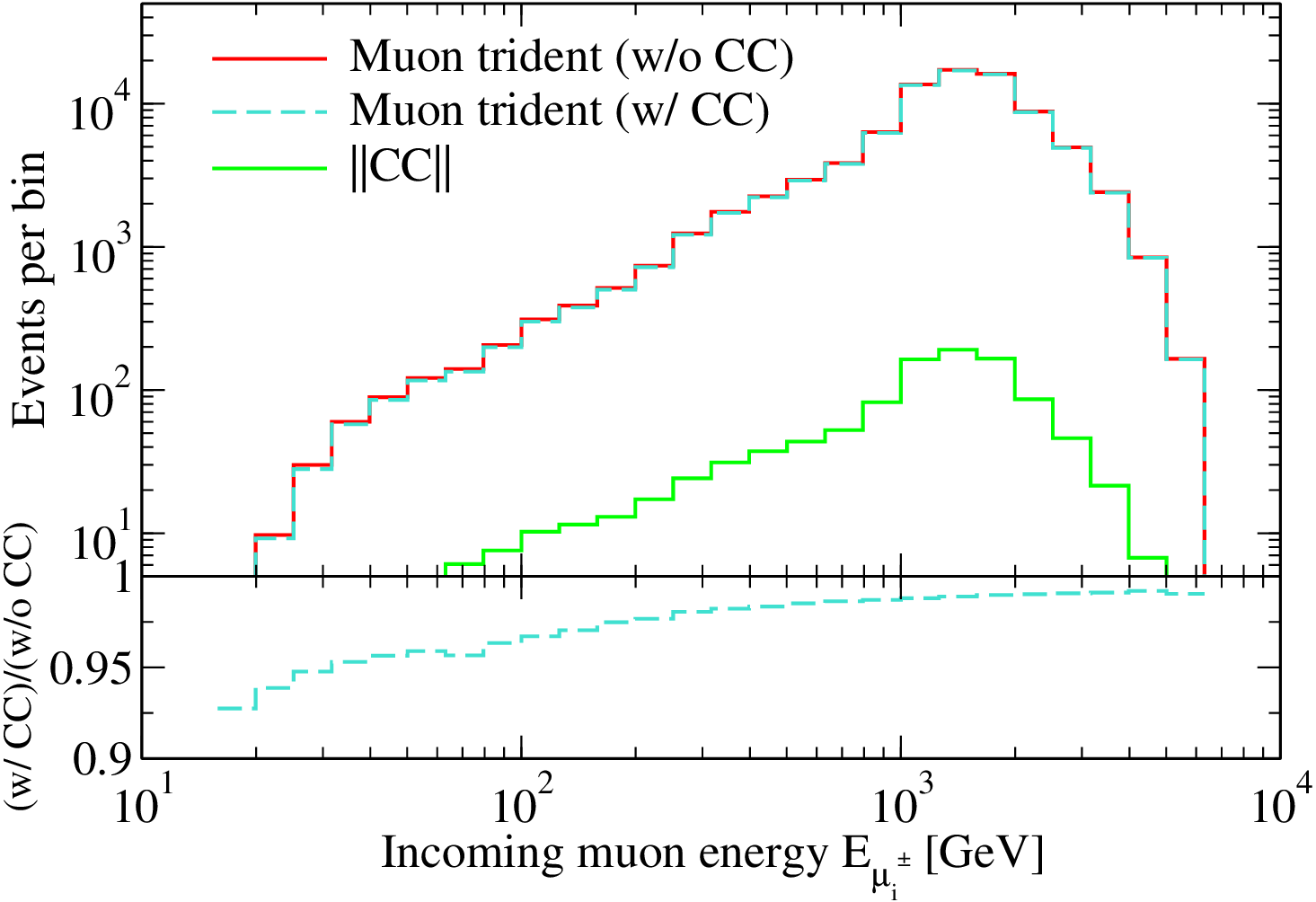}
\end{tabular}
\caption{ Muon trident events binned in the energy of the incident muon for the first Born approximation (red), for the modulus of the Coulomb correction (green) and their sum (turquoise). }
\label{fig:hist_CC}
\end{figure}

In Fig.~\ref{fig:hist_CC} we compare our predictions for events binned in the incident muon energy for coherent scattering. The red line corresponds to our baseline prediction associated with the one-photon exchange, represented by the diagrams of Fig.~\ref{fig:diagrams}. The modulus of the contribution associated with the Coulomb corrections (see Fig.~\ref{fig:diagrams_CC}) is represented by the green line in Fig.~\ref{fig:hist_CC}. We have that the magnitude of the correction depends on the incident muon energy, and as it is negative, its inclusion implies the decreasing of the number of events by $\approx 7\%$ at 100~GeV and by $\approx 1.2\%$ at 1~TeV.

\begin{figure}[t]
	\centering
	\begin{tabular}{c}
\includegraphics[width=0.45\textwidth]{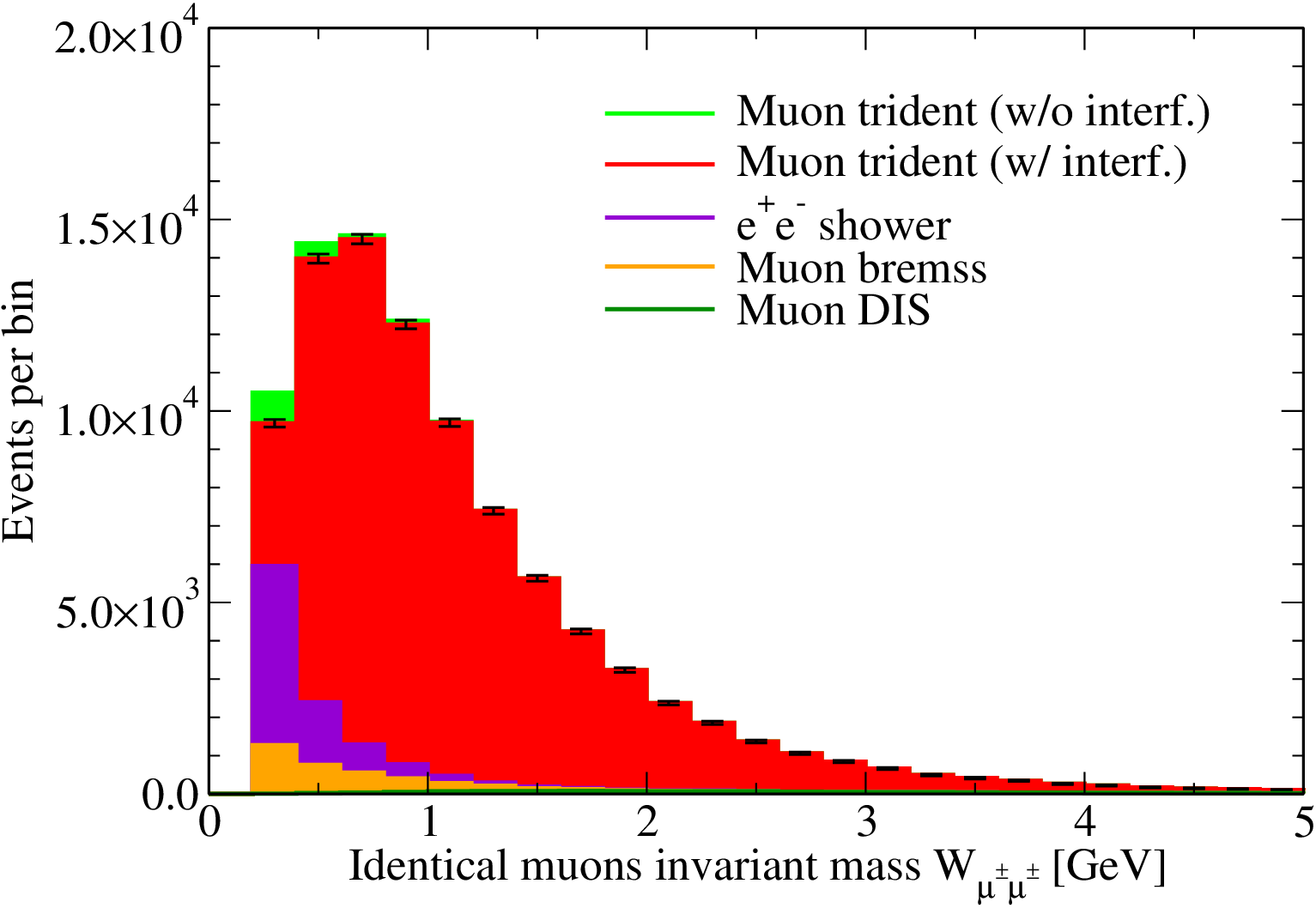}
\end{tabular}
\caption{ Muon trident events at FASER after acceptance cuts binned in the invariant mass of the two identical muons system. We are comparing results with (red) and without (magenta) the interference term from identical muons. 
}
\label{fig:W_identicalMuons}
\end{figure}

Another interesting aspect that can be investigated is the impact of the interference term. The presence of two identical muons in the final state implies the interference between amplitudes involving exchanged muons. This interference contribution is expected to be larger when the identical muons populate similar regions of phase space. We have verified that the impact of the interference contribution is small on the predictions for the total cross-section. In Fig.~\ref{fig:W_identicalMuons} we compare predictions with and without this interference term for events at FASER binned in the invariant mass of the two identical muons' system. Our results indicate that this interference affects mainly small invariant mass region, and becomes negligible for invariant masses above approximately $0.6$~GeV. 

\begin{figure}[t]
	\centering
	\begin{tabular}{c}
\includegraphics[width=0.47\textwidth]{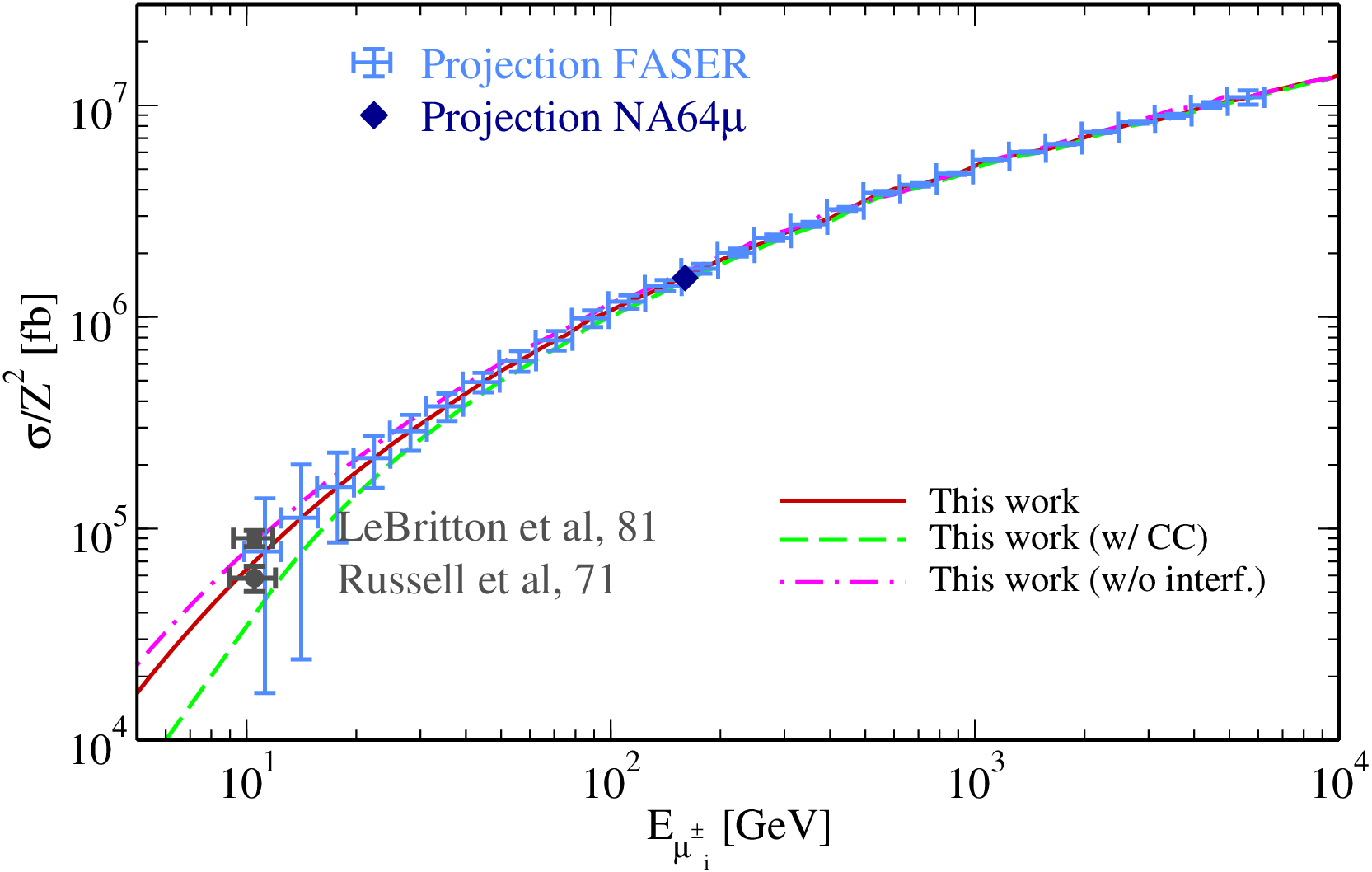} 
	\end{tabular}
\caption{Predictions for the trident cross-section, normalized by $Z^2$, as a function of incident muon energy, derived assuming different assumptions in the calculation. Experimental data from Refs.~\cite{Russell:1971mf,Lebritton:1979gd}. In addition, the predicted sensitivity of the FASER and NA64$\mu$ experiments is presented. 
}
\label{fig:projected_sensitivity_CC}
\end{figure}

Finally, in Fig.~\ref{fig:projected_sensitivity_CC} we compare the muon trident predictions with those derived taking into account the Coulomb corrections (first Born approximation  + CC) and disregarding the interference contribution (first Born approximation w/o interf.). We see that the Coulomb corrections decreases the cross-sections at lower energies. In contrast, if the interference contribution is not considered the magnitude of the cross-section at low energies is increased. However, throughout most of the energy range probed by the FASER and NA64$\mu$ experiments the impact is small.  
In addition, we present in Fig.~\ref{fig:projected_sensitivity_CC} the projected sensitivity of a FASER measurement at the end of Run~3 of the LHC using the tracking system and for the NA64$\mu$ experiment using a muon flux of 160~GeV and assuming $10^{13}$~MOT, which is expected for its phase II. We also show the two existing measurements for this process obtained with 10.5~GeV accelerator muon beams at the Brookhaven National Laboratory~\cite{Russell:1971mf,Lebritton:1979gd}.

\subsection{Impact of electroweak corrections}

In addition to the diagrams associated with the photon exchange, discussed in the previous section, we have also estimated the cross-section including the diagrams where a $Z$ gauge boson is exchanged by the incident muon. We have verified that this correction modifies the expected event rate by less than 0.01\%. Such a result is expected, since  the electroweak (EW) corrections for the muon-nucleus cross-section for the trident process are proportional to ${q^2 g_V^2}/(q^2-m^2_{Z})$, where $q$ is the four-momentum exchanged by the incident muon. At the energy range and invariant mass of the leptons in the final states considered in this study, the cross-section is dominated by values of $q^2 \approx 0.01\, \mathrm{GeV}^2$, which suppress the EW correction (for a related discussion see, e.g., Ref.~\cite{Godunov:2023myj}).

\subsection{Impact of BSM Physics} 

Over the last years, several groups have investigated the potentiality of the neutrino trident process to probe for signals of new physics (see, e.g., Refs.~\cite{Francener:2024jra,Altmannshofer:2019zhy,Altmannshofer:2024hqd,Francener:2026ukt,Kling:2026ghd}). Here, we extend these studies for the case of the muon trident process and perform a first exploratory investigation whether it can provide additional constraints in two alternative BSM scenarios that probe distinct mass ranges of the new particles. In particular, we consider the following scenarios: (a) the $L_\mu - L_\tau$ model \cite{He:1990pn,He:1991qd}, where a new $Z'$ boson is predicted, which couples to muons,  and (b)  the Standard Model Effective Field Theory (SMEFT) \cite{Buchmuller:1985jz}.

We have included the contribution of the associated diagrams with the $Z'$ gauge boson predicted by the $L_\mu - L_\tau$ model to the scattering amplitude of the trident process. To estimate the best possible sensitivity, we consider only statistical uncertainties and the total event rates. We find that a search for deviations in the muon trident cross-section from its theoretical prediction at FASER (NA64$\mu$) is able to probe regions of the coupling constant $g'$ close to $10^{-2}$ ($4\times 10^{-3}$) for $m_{Z'} < 0.1\,\mathrm{GeV}$. For larger masses of the $Z'$ boson, FASER (NA64$\mu$) is able to probe regions with $g'/m_{Z'}$ approximately $7.6\times10^{-2}\, \mathrm{GeV}^{-1}$ ($2.4\times10^{-2}\, \mathrm{GeV}^{-1}$). These bounds are approximately one order of magnitude less stringent than current bounds. This estimate therefore shows that a measurement of the muon trident process is not able to probe any new region of the parameter space of the $Z'$ gauge boson.

In the SMEFT case, a four-lepton contact interaction arises from dimension-six operators in the Warsaw basis, and generates an effective four-muon interaction of the form
\begin{equation}
\mathcal{L}_{\mathrm{eff}} \supset \frac{C_{\mu\mu}}{\Lambda^{2}}(\bar{\mu}_R \gamma^{\mu}\mu_R)(\bar{\mu}_R \gamma_{\mu}\mu_R) \, ,
\label{eq:SMEFT_operator}
\end{equation}
where $C_{\mu\mu}$ is the Wilson coefficient and $\Lambda$ is the EFT scale. We focus on this operator because, unlike operators involving left-handed muon currents, it is not directly constrained by neutrino trident production, making it a particularly interesting probe of four-muon contact interactions. The current bound on this operator is $|C_{\mu \mu}| = 16.9$ for $ \Lambda = 246$~GeV as obtained from the decay $Z \to 4\mu$ at 68\% confidence level~\cite{Boughezal:2020klp}.
Saturating the current bound, we find that the inclusion of the diagrams associated with SMEFT operator implies that the expected number of events is modified by approximately 0.04\% at FASER and at NA64$\mu$ experiments. This weak sensitivity on the SMEFT operator can be understood by the small momentum transfer $Q$ by the incident muon in addition to suppression by the EFT scale $\Lambda$.
 
While the previous discussion focused on the total event rates, both BSM scenarios also affect kinematic distributions. The most interesting candidate is the distribution of the momentum transferred by the incident muon. However, this quantity is not measurable due to the lack of an initial muon momentum measurement at FASER. Other distributions were found to change only mildly or are sensitive only in phase space regions with less than one expected event. Overall, we conclude the muon tridents are not a good laboratory for new physics searches.

\section{Summary} 
\label{sec:sum}

Although quantum electrodynamics is considered the  most rigorously tested and accurate theory in the Standard Model, the theoretical description and experimental probing of the  electromagnetic production of charged muon pairs by an incident muon in the Coulomb field of a heavy nucleus - the muon trident process - are still in an embryonic phase. Currently, the  model predictions for the behavior of the associated cross-section   
at TeV energies differ significantly, and a limited set of experimental data is available. Such aspects have motivated the analysis performed in this work.

We have investigated the prospects for measuring the muon trident process with high-energy collider muons at the FASER experiment during Run~3 of the LHC. Our analysis was performed using a full tree-level calculation of the trident cross-section, including all exchange diagrams associated with identical muons in the final state, precise nuclear form factors, and acceptance cuts from the FASER detector geometry and tracking system. Our main result is that during LHC Run~3 the FASER experiment is expected to observe more than $8.6\times10^4$   muon trident events after acceptance cuts, allowing for the first measurement of this process with TeV muons and with percent-level statistical precision. 

Several aspects have been investigated, and the main conclusions can be summarized as follows: (a) We have estimated the contribution of coherent and incoherent interactions for the muon trident process and found that the incoherent contribution is strongly suppressed after detector acceptance requirements, allowing us to constrain the theoretical description of the coherent trident process; (b) The dominant backgrounds associated with muon bremsstrahlung, electromagnetic showers from $e^+e^-$ production, and muon DIS with hadrons misidentified as muons in the final state were estimated, and our results indicate that these backgrounds remain subleading compared to the trident signal and may be further reduced through dedicated kinematic selections; (c) The muon trident process is strongly sensitive to the modeling of nuclear form factors, with  simplified parametrizations overestimating the total cross-section by more than $5\%$ at TeV energies when compared to predictions based on Woods-Saxon charge distributions; (d) The impact of the Coulomb corrections was estimated for the first time, with the results demonstrating that these corrections are dependent of the incident muon energy and reduce the trident event rate by approximately $1.4\%$ at FASER; (e) The contribution of the interference terms associated with identical muons in the final state was analyzed,  and the results indicate that the impact of the interference term is larger for  low values of the invariant mass of the identical muon system;  (f) The impact of electroweak corrections associated with the $Z$ boson exchange was estimated, with the impact being small in the kinematical range considered;  (g) Two scenarios for BSM physics  ($L_\mu - L_\tau$  and SMEFT) were considered and its impact on the trident cross-section was estimated assuming a set of parameters in an unconstrained region, with the results indicating  a small contribution from the new physics diagrams; and (h) Predictions for the NA64$\mu$ experiment have also been presented, which indicate that this experiment is expected to observe $8.8\times10^7$  muon trident events.

The detailed analysis performed in this paper indicates that the investigation of the  muon trident scattering with high-energy muons is feasible  and that a future experimental study will be a promising laboratory to investigate QED effects in the Coulomb field of a heavy nucleus.

\section*{Data Availability}

The results of this work can be reproduced using the Monte Carlo generator available in: \url{https://github.com/reinaldofrancener/MuonTridents}

\section*{Acknowledgments}

The authors would like to thank Toni Mäkelä for providing the EM shower simulation.  V. P. G. would like to thank the
members of the Institute of Modern Physics, Chinese Academy of Sciences, Huizhou, China, for their warm hospitality
during the completion of this study. The work of R.F. was supported by FAPERGS. V.P.G. was partially financed by the Brazilian funding agencies CNPq, FAPERGS and INCT-FNA  (Process No. 408419/2024-5). The work of G.R.S. was supported by the CAPES doctoral fellowship 88887.005836/2024-00.

\appendix

\section{ Form factors and nuclear charge distributions}
\label{appendix:FormFactor}
The nuclear form factor is defined by the Fourier transform of the charge distribution, 
\begin{equation}
    F(\mathbf{q}^{2}) =  \int d^{3} r \rho(r) e^{-i \mathbf{q} \cdot\mathbf{r}}
\end{equation}
For a  spherically symmetric charge distribution, one has 
\begin{equation}
    F(\mathbf{q}) = \frac{4 \pi}{q} \int^{\infty}_0 dr \, r \rho(r) \sin(r q).
    \label{eq:FFexpr}
\end{equation}
In what follows,  we will present the form factors employed in our analysis that are described by an  analytical expression. It is important to emphasize that in our analysis we also have considered the point-like form factor, in which the charge distribution is considered to be located in one single point and $F(\mathbf{q}) = 1$.

\subsection{Monopole Form Factor}
 The charge distribution $\rho(r)$ of  the nuclei can be described by a Yukawa distribution~\cite{Hencken:1993cf}, which is given by
\begin{equation}
    \rho(r) =  \frac{\Lambda^{2}}{4 \pi} \frac{e^{- \Lambda r}}{r},
    \label{eq:yukdist}
\end{equation}
where $\Lambda$ is a constant that in principle can be  constrained  experimentally and that  is related to the radius of the nuclei. The corresponding form factor is given by 
\begin{equation}
    \begin{aligned}
        F(\mathbf{q}) &= \frac{\Lambda^{2}}{q} \int^{\infty}_0 dr \, e^{- \Lambda r} \sin(r q) = \frac{\Lambda^{2}}{\Lambda^{2} + q^{2}}.
    \end{aligned}
    \label{eq:FFmonopole}
\end{equation}

\subsection{Woods-Saxon and Realistic Form Factors}
The charge distribution in nuclei is more precisely described using the two-parameter Fermi model (Woods-Saxon distribution), given by
\begin{equation}
   \rho(r) = \frac{\rho_0}{1 + \exp[(r- R)/a]} ,
\end{equation}
with  $a$ being the skin depth and  $R$ the nuclear radius. However, the  calculation of the form factor associated with this distribution is not a simple task and requires a numerical implementation. An alternative can be derived considering  the convolution between a constant charge distribution in the nuclei, $\rho(r) = \rho_0/A$ ($A$ being the mass number) and the monopole form factor Eq.~(\ref{eq:FFmonopole}). Such form factor is usually denoted realistic form factor and is given by
\begin{equation}
    F(\mathbf{q}) = \frac{4 \pi \rho_0}{q^{3} A} \left[ \sin(q R) - qR \cos(qR) \right]  \frac{\Lambda^{2}}{\Lambda^{2} + q^{2}}.
\end{equation}

\bibliographystyle{utphys}
\bibliography{ref}  

\providecommand{\href}[2]{#2}\begingroup\raggedright\begin{thebibliography}{10}

\bibitem{Morris1968}
M.~L. Morris and R.~O. Stenerson, ``Muon-pair production by cosmic-ray muons,''
  \href{http://dx.doi.org/10.1007/BF02710250}{{\em Il Nuovo Cimento B
  (1965-1970)} {\bf 53} (1968) no.~2, 494--506}.
  \url{https://doi.org/10.1007/BF02710250}.

\bibitem{Barton1970}
J.~C. Barton and I.~W. Rogers, ``Proceedings contribution in the 11th
  international conference on cosmic rays,'' in {\em Proceedings of the 11th
  International Conference on Cosmic Rays}, A.~Somogyi, ed., vol.~29, p.~259.
\newblock Budapest, Hungary, 1970.

\bibitem{Russell:1971mf}
J.~J. Russell, R.~C. Sah, M.~J. Tannenbaum, W.~E. Cleland, D.~G. Ryan, and
  D.~G. Stairs, ``{Observation of muon trident production in lead and the
  statistics of the muon},''
  \href{http://dx.doi.org/10.1103/PhysRevLett.26.46}{{\em Phys. Rev. Lett.}
  {\bf 26} (1971)  46--50}.

\bibitem{Lebritton:1979gd}
J.~Lebritton {\em et al.}, ``{Trimuon and Dimuon Events Produced by
  10.5-{GeV}/$c \mu^-$ on Lead},''
  \href{http://dx.doi.org/10.1103/PhysRevD.23.1}{{\em Phys. Rev. D} {\bf 23}
  (1981)  1}.

\bibitem{MACRO:1999qxo}
{\bf MACRO} Collaboration, M.~Ambrosio {\em et al.}, ``{High statistics
  measurement of the underground muon pair separation at Gran Sasso},''
  \href{http://dx.doi.org/10.1103/PhysRevD.60.032001}{{\em Phys. Rev. D} {\bf
  60} (1999)  032001}, \href{http://arxiv.org/abs/hep-ex/9901027}{{\tt
  arXiv:hep-ex/9901027}}.

\bibitem{Battistoni:409814}
{\bf MACRO} Collaboration, G.~Battistoni and E.~Scapparone, ``{Hints of $\mu +
  N \to \mu + N + \mu^{+} \mu^{-}$ process observation with the MACRO
  experiment},''
  \href{http://dx.doi.org/https://doi.org/10.1016/S0920-5632(99)00303-5}{{\em
  Nuclear Physics B - P. Supplements} {\bf 75} (1999) no.~1, 394--396}.

\bibitem{Anikeev:2005td}
V.~B. Anikeev {\em et al.}, ``{Direct observation of muon-pair production by
  high-energy muons in the liquid-argon calorimeter BARS},''
  \href{http://dx.doi.org/10.1134/1.1866380}{{\em Phys. Atom. Nucl.} {\bf 68}
  (2005)  259--266}.

\bibitem{Maciuc:2006xb}
F.~Maciuc {\em et al.}, ``{Muon-pair production by atmospheric muons in
  cosmoALEPH},'' \href{http://dx.doi.org/10.1103/PhysRevLett.96.021801}{{\em
  Phys. Rev. Lett.} {\bf 96} (2006)  021801}.

\bibitem{Henry:1967jm}
G.~R. Henry, ``{Trident Production with Nuclear Targets},''
  \href{http://dx.doi.org/10.1103/PhysRev.154.1534}{{\em Phys. Rev.} {\bf 154}
  (1967)  1534--1536}.

\bibitem{Tannenbaum:1968zz}
M.~J. Tannenbaum, ``{Muon Tridents},''
  \href{http://dx.doi.org/10.1103/PhysRev.167.1308}{{\em Phys. Rev.} {\bf 167}
  (1968)  1308--1313}.

\bibitem{Bugaev1970}
E.~V. Bugaev, Y.~D. Kotov, and I.~L. Rozental', {\em Cosmic Muons and
  Neutrinos}.
\newblock Atomizdat, Moscow, 1970.
\newblock p. 38, in Russian.

\bibitem{Barger:1979eg}
V.~D. Barger, W.-Y. Keung, and R.~J.~N. Phillips, ``{MULTI - MUON PRODUCTION BY
  MUONS},'' \href{http://dx.doi.org/10.1103/PhysRevD.20.630}{{\em Phys. Rev. D}
  {\bf 20} (1979)  630}.

\bibitem{Ganapathi:1980wh}
V.~Ganapathi and J.~Smith, ``{Electromagnetic Production of Trimuons in Muon
  Scattering. 3. Bethe-Heitler Reactions With Muon and Heavy Lepton Pairs},''
  \href{http://dx.doi.org/10.1103/PhysRevD.23.75}{{\em Phys. Rev. D} {\bf 23}
  (1981)  75}.

\bibitem{Kelner:1998mh}
S.~R. Kelner, ``{Pair production in collisions between muons and atomic
  electrons},'' {\em Phys. Atom. Nucl.} {\bf 61} (1998)  448--456.

\bibitem{Kelner:2000va}
S.~R. Kelner, R.~P. Kokoulin, and A.~A. Petrukhin, ``{Direct production of muon
  pairs by high-energy muons},''
  \href{http://dx.doi.org/10.1134/1.1312894}{{\em Phys. Atom. Nucl.} {\bf 63}
  (2000)  1603--1611}.

\bibitem{Das:2026eyy}
S.~Das, L.~Kupari, M.~H. Reno, and S.~Trojanowski, ``{Muon-induced di-tau
  production as a probe of new physics},''
  \href{http://arxiv.org/abs/2606.05406}{{\tt arXiv:2606.05406 [hep-ph]}}.

\bibitem{Budnev:1975poe}
V.~M. Budnev, I.~F. Ginzburg, G.~V. Meledin, and V.~G. Serbo, ``{The Two photon
  particle production mechanism. Physical problems. Applications. Equivalent
  photon approximation},''
  \href{http://dx.doi.org/10.1016/0370-1573(75)90009-5}{{\em Phys. Rept.} {\bf
  15} (1975)  181--281}.

\bibitem{FASER:2022hcn}
{\bf FASER} Collaboration, H.~Abreu {\em et al.}, ``{The FASER detector},''
  \href{http://dx.doi.org/10.1088/1748-0221/19/05/P05066}{{\em JINST} {\bf 19}
  (2024) no.~05, P05066}, \href{http://arxiv.org/abs/2207.11427}{{\tt
  arXiv:2207.11427 [physics.ins-det]}}.

\bibitem{SNDLHC:2022ihg}
{\bf SND@LHC} Collaboration, G.~Acampora {\em et al.}, ``{SND@LHC: the
  scattering and neutrino detector at the LHC},''
  \href{http://dx.doi.org/10.1088/1748-0221/19/05/P05067}{{\em JINST} {\bf 19}
  (2024) no.~05, P05067}, \href{http://arxiv.org/abs/2210.02784}{{\tt
  arXiv:2210.02784 [hep-ex]}}.

\bibitem{Feng:2017uoz}
J.~L. Feng, I.~Galon, F.~Kling, and S.~Trojanowski, ``{ForwArd Search
  ExpeRiment at the LHC},''
  \href{http://dx.doi.org/10.1103/PhysRevD.97.035001}{{\em Phys. Rev. D} {\bf
  97} (2018) no.~3, 035001}, \href{http://arxiv.org/abs/1708.09389}{{\tt
  arXiv:1708.09389 [hep-ph]}}.

\bibitem{FASER:2019dxq}
{\bf FASER} Collaboration, H.~Abreu {\em et al.}, ``{Detecting and Studying
  High-Energy Collider Neutrinos with FASER at the LHC},''
  \href{http://dx.doi.org/10.1140/epjc/s10052-020-7631-5}{{\em Eur. Phys. J. C}
  {\bf 80} (2020) no.~1, 61}, \href{http://arxiv.org/abs/1908.02310}{{\tt
  arXiv:1908.02310 [hep-ex]}}.

\bibitem{FASER:2021mtu}
{\bf FASER} Collaboration, H.~Abreu {\em et al.}, ``{First neutrino interaction
  candidates at the LHC},''
  \href{http://dx.doi.org/10.1103/PhysRevD.104.L091101}{{\em Phys. Rev. D} {\bf
  104} (2021) no.~9, L091101}, \href{http://arxiv.org/abs/2105.06197}{{\tt
  arXiv:2105.06197 [hep-ex]}}.

\bibitem{SNDLHC:2023mib}
{\bf SND@LHC} Collaboration, R.~Albanese {\em et al.}, ``{Measurement of the
  muon flux at the SND@LHC experiment},''
  \href{http://dx.doi.org/10.1140/epjc/s10052-023-12380-3}{{\em Eur. Phys. J.
  C} {\bf 84} (2024) no.~1, 90}, \href{http://arxiv.org/abs/2310.05536}{{\tt
  arXiv:2310.05536 [hep-ex]}}.

\bibitem{SNDLHC:2026mip}
{\bf SND@LHC} Collaboration, D.~Abbaneo {\em et al.}, ``{Measurement of the
  Muon Flux at SND@LHC: Results from the 2023-2025 Proton and Heavy-Ion
  Periods},'' \href{http://arxiv.org/abs/2602.23412}{{\tt arXiv:2602.23412
  [hep-ex]}}.

\bibitem{Francener:2025pnr}
R.~Francener, V.~P. Goncalves, F.~Kling, P.~Krack, and J.~Rojo,
  ``{Deep-inelastic scattering at TeV energies with LHC muons},''
  \href{http://dx.doi.org/10.1140/epjc/s10052-025-14829-z}{{\em Eur. Phys. J.
  C} {\bf 85} (2025) no.~10, 1098}, \href{http://arxiv.org/abs/2506.13889}{{\tt
  arXiv:2506.13889 [hep-ph]}}.

\bibitem{Francener:2025tyh}
R.~Francener, V.~P. Goncalves, and D.~R. Gratieri, ``{Investigating nuclear
  effects in lepton-ion DIS at the LHC},''
  \href{http://dx.doi.org/10.1007/JHEP01(2026)149}{{\em JHEP} {\bf 01} (2026)
  149}, \href{http://arxiv.org/abs/2509.00144}{{\tt arXiv:2509.00144
  [hep-ph]}}.

\bibitem{Hernandez-Sainz:2026rxc}
P.~Hernandez-Sainz, F.~Kling, J.~Koorn, and J.~Rojo, ``{Event Generation for
  Neutrino and Muon DIS at the LHC},''
  \href{http://arxiv.org/abs/2610.16836}{{\tt arXiv:2610.16836 [hep-ph]}}.

\bibitem{Ariga:2023fjg}
A.~Ariga, R.~Balkin, I.~Galon, E.~Kajomovitz, and Y.~Soreq, ``{Hunting muonic
  forces at emulsion detectors},''
  \href{http://dx.doi.org/10.1103/PhysRevD.109.035003}{{\em Phys. Rev. D} {\bf
  109} (2024) no.~3, 035003}, \href{http://arxiv.org/abs/2305.03102}{{\tt
  arXiv:2305.03102 [hep-ph]}}.

\bibitem{Batell:2024cdl}
B.~Batell, H.~Davoudiasl, R.~Marcarelli, E.~T. Neil, and S.~Trojanowski,
  ``{Lepton-flavor-violating ALP signals with TeV-scale muon beams},''
  \href{http://dx.doi.org/10.1103/PhysRevD.110.075039}{{\em Phys. Rev. D} {\bf
  110} (2024) no.~7, 075039}, \href{http://arxiv.org/abs/2407.15942}{{\tt
  arXiv:2407.15942 [hep-ph]}}.

\bibitem{MammenAbraham:2025gai}
R.~Mammen~Abraham and M.~Fieg, ``{LHC as a TeV muon beam dump: Muonphilic
  scalars at FASER},'' \href{http://dx.doi.org/10.1103/11j5-txpc}{{\em Phys.
  Rev. D} {\bf 112} (2025) no.~1, 015005},
  \href{http://arxiv.org/abs/2501.09071}{{\tt arXiv:2501.09071 [hep-ph]}}.

\bibitem{Francener:2025wzh}
R.~Francener, V.~P. Goncalves, and G.~Rabelo-Soares, ``{Muon trident process at
  far-forward LHC detectors},''
  \href{http://dx.doi.org/10.1016/j.nuclphysb.2026.117396}{{\em Nucl. Phys. B}
  {\bf 1025} (2026)  117396}, \href{http://arxiv.org/abs/2510.18943}{{\tt
  arXiv:2510.18943 [hep-ph]}}.

\bibitem{SNDLHC:2024bzp}
{\bf SND@LHC} Collaboration, D.~Abbaneo {\em et al.}, ``{Results and
  Perspectives from the First Two Years of Neutrino Physics at the LHC by the
  SND@LHC Experiment},'' \href{http://dx.doi.org/10.3390/sym16060702}{{\em
  Symmetry} {\bf 16} (2024) no.~6, 702}.

\bibitem{Guler:2026mrt}
{\bf SND@LHC} Collaboration, A.~M. G{\"u}ler and O.~Durhan, ``{Search for
  muonic trident production in SND@LHC},''
  \href{http://dx.doi.org/10.22323/1.485.0183}{{\em PoS} {\bf EPS-HEP2025}
  (2026)  183}.

\bibitem{Davies:1954zz}
H.~Davies, H.~A. Bethe, and L.~C. Maximon, ``{Theory of Bremsstrahlung and Pair
  Production. 2. Integral Cross Section for Pair Production},''
  \href{http://dx.doi.org/10.1103/PhysRev.93.788}{{\em Phys. Rev.} {\bf 93}
  (1954)  788--795}.

\bibitem{Baltz:2009fs}
A.~J. Baltz, ``{Higher Order QED Calculation of Ultrarelativistic Heavy Ion
  Production of mu+ mu- Pairs},''
  \href{http://dx.doi.org/10.1103/PhysRevC.80.034901}{{\em Phys. Rev. C} {\bf
  80} (2009)  034901}, \href{http://arxiv.org/abs/0901.0891}{{\tt
  arXiv:0901.0891 [nucl-th]}}.

\bibitem{Jentschura:2009mb}
U.~D. Jentschura and V.~G. Serbo, ``{Nuclear form factor, validity of the
  equivalent photon approximation and Coulomb corrections to muon pair
  production in photon-nucleus and nucleus-nucleus collisions},''
  \href{http://dx.doi.org/10.1140/epjc/s10052-009-1147-3}{{\em Eur. Phys. J. C}
  {\bf 64} (2009)  309--317}, \href{http://arxiv.org/abs/0908.3853}{{\tt
  arXiv:0908.3853 [hep-ph]}}.

\bibitem{Li:2023yjt}
X.~Li, J.~Luo, Z.~Tang, X.~Wu, and W.~Zha, ``{Exploring the higher-order QED
  effects on the differential distributions of vacuum pair production in
  relativistic heavy-ion collisions},''
  \href{http://dx.doi.org/10.1016/j.physletb.2023.138314}{{\em Phys. Lett. B}
  {\bf 847} (2023)  138314}, \href{http://arxiv.org/abs/2307.01549}{{\tt
  arXiv:2307.01549 [hep-ph]}}.

\bibitem{GEANT4:2002zbu}
{\bf GEANT4} Collaboration, S.~Agostinelli {\em et al.}, ``{GEANT4 - A
  Simulation Toolkit},''
  \href{http://dx.doi.org/10.1016/S0168-9002(03)01368-8}{{\em Nucl. Instrum.
  Meth. A} {\bf 506} (2003)  250--303}.

\bibitem{Yajaman2022}
S.~Yajaman, ``The implementation of {$\mu^- \mu^+$} production by muons in
  geant4,'' Tech. Rep. CERN-STUDENTS-Note-2022-008, CERN, Geneva, 2022.
\newblock \url{https://repository.cern/records/8afar-vxz20}.
\newblock CERN Summer Student Programme.

\bibitem{Koehne:2013gpa}
J.~H. Koehne, K.~Frantzen, M.~Schmitz, T.~Fuchs, W.~Rhode, D.~Chirkin, and
  J.~Becker~Tjus, ``{PROPOSAL: A tool for propagation of charged leptons},''
  \href{http://dx.doi.org/10.1016/j.cpc.2013.04.001}{{\em Comput. Phys.
  Commun.} {\bf 184} (2013)  2070--2090}.

\bibitem{Dunsch:2018nsc}
M.~Dunsch, J.~Soedingrekso, A.~Sandrock, M.~Meier, T.~Menne, and W.~Rhode,
  ``{Recent Improvements for the Lepton Propagator PROPOSAL},''
  \href{http://dx.doi.org/10.1016/j.cpc.2019.03.021}{{\em Comput. Phys.
  Commun.} {\bf 242} (2019)  132--144},
  \href{http://arxiv.org/abs/1809.07740}{{\tt arXiv:1809.07740 [hep-ph]}}.

\bibitem{Albright:1978mg}
C.~H. Albright, J.~Smith, and J.~A.~M. Vermaseren, ``{A Comparison of Trimuon
  Production Mechanisms},''
  \href{http://dx.doi.org/10.1103/PhysRevD.18.108}{{\em Phys. Rev. D} {\bf 18}
  (1978)  108}.

\bibitem{Ganapathi:1978qm}
V.~Ganapathi and J.~Smith, ``{Electromagnetic Production of Trimuons in Deep
  Inelastic Muon Scattering},''
  \href{http://dx.doi.org/10.1103/PhysRevD.19.801}{{\em Phys. Rev. D} {\bf 19}
  (1979)  801--809}.

\bibitem{vonWeizsacker:1934nji}
C.~F. von Weizsacker, ``{Radiation emitted in collisions of very fast
  electrons},'' \href{http://dx.doi.org/10.1007/BF01333110}{{\em Z. Phys.} {\bf
  88} (1934)  612--625}.

\bibitem{Williams:1934ad}
E.~J. Williams, ``{Nature of the high-energy particles of penetrating radiation
  and status of ionization and radiation formulae},''
  \href{http://dx.doi.org/10.1103/PhysRev.45.729}{{\em Phys. Rev.} {\bf 45}
  (1934)  729--730}.

\bibitem{Bertulani:1987tz}
C.~A. Bertulani and G.~Baur, ``{Electromagnetic Processes in Relativistic Heavy
  Ion Collisions},'' \href{http://dx.doi.org/10.1016/0370-1573(88)90142-1}{{\em
  Phys. Rept.} {\bf 163} (1988)  299}.

\bibitem{Breit:1934zz}
G.~Breit and J.~A. Wheeler, ``{Collision of two light quanta},''
  \href{http://dx.doi.org/10.1103/PhysRev.46.1087}{{\em Phys. Rev.} {\bf 46}
  (1934) no.~12, 1087--1091}.

\bibitem{Mertig:1990an}
R.~Mertig, M.~Bohm, and A.~Denner, ``{FEYN CALC: Computer algebraic calculation
  of Feynman amplitudes},''
  \href{http://dx.doi.org/10.1016/0010-4655(91)90130-D}{{\em Comput. Phys.
  Commun.} {\bf 64} (1991)  345--359}.

\bibitem{Shtabovenko:2016sxi}
V.~Shtabovenko, R.~Mertig, and F.~Orellana, ``{New Developments in FeynCalc
  9.0},'' \href{http://dx.doi.org/10.1016/j.cpc.2016.06.008}{{\em Comput. Phys.
  Commun.} {\bf 207} (2016)  432--444},
  \href{http://arxiv.org/abs/1601.01167}{{\tt arXiv:1601.01167 [hep-ph]}}.

\bibitem{Shtabovenko:2020gxv}
V.~Shtabovenko, R.~Mertig, and F.~Orellana, ``{FeynCalc 9.3: New features and
  improvements},'' \href{http://dx.doi.org/10.1016/j.cpc.2020.107478}{{\em
  Comput. Phys. Commun.} {\bf 256} (2020)  107478},
  \href{http://arxiv.org/abs/2001.04407}{{\tt arXiv:2001.04407 [hep-ph]}}.

\bibitem{Altmannshofer:2019zhy}
W.~Altmannshofer, S.~Gori, J.~Mart{\'\i}n-Albo, A.~Sousa, and M.~Wallbank,
  ``{Neutrino Tridents at DUNE},''
  \href{http://dx.doi.org/10.1103/PhysRevD.100.115029}{{\em Phys. Rev. D} {\bf
  100} (2019) no.~11, 115029}, \href{http://arxiv.org/abs/1902.06765}{{\tt
  arXiv:1902.06765 [hep-ph]}}.

\bibitem{Woods:1954zz}
R.~D. Woods and D.~S. Saxon, ``{Diffuse Surface Optical Model for
  Nucleon-Nuclei Scattering},''
  \href{http://dx.doi.org/10.1103/PhysRev.95.577}{{\em Phys. Rev.} {\bf 95}
  (1954)  577--578}.

\bibitem{DeVries:1987atn}
H.~De~Vries, C.~W. De~Jager, and C.~De~Vries, ``{Nuclear charge and
  magnetization density distribution parameters from elastic electron
  scattering},'' \href{http://dx.doi.org/10.1016/0092-640X(87)90013-1}{{\em
  Atom. Data Nucl. Data Tabl.} {\bf 36} (1987)  495--536}.

\bibitem{Bell:1963ogq}
J.~S. Bell and M.~J.~G. Veltman, ``{Intermediate boson production by
  neutrinos},'' \href{http://dx.doi.org/10.1016/S0375-9601(63)80045-6}{{\em
  Phys. Lett.} {\bf 5} (1963)  94--96}.

\bibitem{Yukawa:1935xg}
H.~Yukawa, ``{On the Interaction of Elementary Particles I},''
  \href{http://dx.doi.org/10.1143/PTPS.1.1}{{\em Proc. Phys. Math. Soc. Jap.}
  {\bf 17} (1935)  48--57}.

\bibitem{Klein:1999qj}
S.~Klein and J.~Nystrand, ``{Exclusive vector meson production in relativistic
  heavy ion collisions},''
  \href{http://dx.doi.org/10.1103/PhysRevC.60.014903}{{\em Phys. Rev. C} {\bf
  60} (1999)  014903}, \href{http://arxiv.org/abs/hep-ph/9902259}{{\tt
  arXiv:hep-ph/9902259}}.

\bibitem{Schiff:1951zza}
L.~I. Schiff, ``{Energy-Angle Distribution of Thin Target Bremsstrahlung},''
  \href{http://dx.doi.org/10.1103/PhysRev.83.252}{{\em Phys. Rev.} {\bf 83}
  (1951)  252--253}.

\bibitem{Tsai:1973py}
Y.-S. Tsai, ``{Pair Production and Bremsstrahlung of Charged Leptons},''
  \href{http://dx.doi.org/10.1103/RevModPhys.46.815}{{\em Rev. Mod. Phys.} {\bf
  46} (1974)  815}. [Erratum: Rev.Mod.Phys. 49, 421--423 (1977)].

\bibitem{FASER:2018bac}
{\bf FASER} Collaboration, A.~Ariga {\em et al.}, ``{Technical Proposal for
  FASER: ForwArd Search ExpeRiment at the LHC},''
  \href{http://arxiv.org/abs/1812.09139}{{\tt arXiv:1812.09139
  [physics.ins-det]}}.

\bibitem{FASER:2020gpr}
{\bf FASER} Collaboration, H.~Abreu {\em et al.}, ``{Technical Proposal:
  FASERnu},'' \href{http://arxiv.org/abs/2001.03073}{{\tt arXiv:2001.03073
  [physics.ins-det]}}.

\bibitem{FASER:2021ljd}
{\bf FASER} Collaboration, H.~Abreu {\em et al.}, ``{The tracking detector of
  the FASER experiment},''
  \href{http://dx.doi.org/10.1016/j.nima.2022.166825}{{\em Nucl. Instrum. Meth.
  A} {\bf 1034} (2022)  166825}, \href{http://arxiv.org/abs/2112.01116}{{\tt
  arXiv:2112.01116 [physics.ins-det]}}.

\bibitem{FASER:2025qaf}
{\bf FASER} Collaboration, R.~Mammen~Abraham {\em et al.}, ``{Reconstruction
  and performance evaluation of FASER's emulsion detector at the LHC},''
  \href{http://dx.doi.org/10.1088/1748-0221/20/12/P12018}{{\em JINST} {\bf 20}
  (2025) no.~12, P12018}, \href{http://arxiv.org/abs/2504.13008}{{\tt
  arXiv:2504.13008 [physics.ins-det]}}.

\bibitem{FASER:2026hzm}
{\bf FASER} Collaboration, R.~Mammen~Abraham {\em et al.}, ``{Momentum
  Measurement of Charged Particles in FASER's Emulsion Detector at the LHC},''
  \href{http://arxiv.org/abs/2602.17575}{{\tt arXiv:2602.17575 [hep-ex]}}.

\bibitem{Battistoni:2015epi}
G.~Battistoni {\em et al.}, ``{Overview of the FLUKA code},''
  \href{http://dx.doi.org/10.1016/j.anucene.2014.11.007}{{\em Annals Nucl.
  Energy} {\bf 82} (2015)  10--18}.

\bibitem{Sabate-Gilarte:2023aeg}
M.~Sabate-Gilarte and F.~Cerutti, ``{Radiation field characterization for
  present LHC and future HL-LHC forward physics experiments},''
  \href{http://dx.doi.org/10.18429/JACoW-IPAC2023-MOPL018}{{\em JACoW} {\bf
  IPAC2023} (2023)  MOPL018}.

\bibitem{Buckley:2014ana}
A.~Buckley, J.~Ferrando, S.~Lloyd, K.~Nordstr{\"o}m, B.~Page, M.~R{\"u}fenacht,
  M.~Sch{\"o}nherr, and G.~Watt, ``{LHAPDF6: parton density access in the LHC
  precision era},''
  \href{http://dx.doi.org/10.1140/epjc/s10052-015-3318-8}{{\em Eur. Phys. J. C}
  {\bf 75} (2015)  132}, \href{http://arxiv.org/abs/1412.7420}{{\tt
  arXiv:1412.7420 [hep-ph]}}.

\bibitem{Gertsenberger:2023ree}
S.~V. Gertsenberger, ``{THE NA64$\mu$ Experiment at the CERN SPS},''
  \href{http://dx.doi.org/10.1134/S1063778823060133}{{\em Phys. Atom. Nucl.}
  {\bf 86} (2023) no.~6, 1310--1314}.

\bibitem{Chen:2018vkr}
C.-Y. Chen, J.~Kozaczuk, and Y.-M. Zhong, ``{Exploring leptophilic dark matter
  with NA64-$\mu$},'' \href{http://dx.doi.org/10.1007/JHEP10(2018)154}{{\em
  JHEP} {\bf 10} (2018)  154}, \href{http://arxiv.org/abs/1807.03790}{{\tt
  arXiv:1807.03790 [hep-ph]}}.

\bibitem{NA64:2018iqr}
{\bf NA64} Collaboration, ``{Addendum to the Proposal P348: Search for dark
  sector particles weakly coupled to muon with NA64 {\ensuremath{\mu}}},''.

\bibitem{NA64:2024nwj}
{\bf NA64} Collaboration, Y.~M. Andreev {\em et al.}, ``{Shedding light on dark
  sectors with high-energy muons at the NA64 experiment at the CERN SPS},''
  \href{http://dx.doi.org/10.1103/PhysRevD.110.112015}{{\em Phys. Rev. D} {\bf
  110} (2024) no.~11, 112015}, \href{http://arxiv.org/abs/2409.10128}{{\tt
  arXiv:2409.10128 [hep-ex]}}.

\bibitem{Kling:2025lnt}
F.~Kling, T.~M{\"a}kel{\"a}, and J.~McFayden, ``{Antielectron neutrinos at
  high-energy neutrino experiments: Identification strategies and physics
  potential},'' \href{http://dx.doi.org/10.1103/9757-h3f6}{{\em Phys. Rev. D}
  {\bf 113} (2026) no.~1, 013004}, \href{http://arxiv.org/abs/2507.00262}{{\tt
  arXiv:2507.00262 [hep-ph]}}.

\bibitem{Banfi:2023mhz}
A.~Banfi, S.~Ferrario~Ravasio, B.~J{\"a}ger, A.~Karlberg, F.~Reichenbach, and
  G.~Zanderighi, ``{A POWHEG generator for deep inelastic scattering},''
  \href{http://dx.doi.org/10.1007/JHEP02(2024)023}{{\em JHEP} {\bf 02} (2024)
  023}, \href{http://arxiv.org/abs/2309.02127}{{\tt arXiv:2309.02127
  [hep-ph]}}.

\bibitem{vanBeekveld:2024ziz}
M.~van Beekveld, S.~Ferrario~Ravasio, E.~Groenendijk, P.~Krack, J.~Rojo, and
  V.~S. S{\'a}nchez, ``{A phenomenological analysis of LHC neutrino scattering
  at NLO accuracy matched to parton showers},''
  \href{http://dx.doi.org/10.1140/epjc/s10052-024-13386-1}{{\em Eur. Phys. J.
  C} {\bf 84} (2024) no.~11, 1175}, \href{http://arxiv.org/abs/2407.09611}{{\tt
  arXiv:2407.09611 [hep-ph]}}.

\bibitem{FerrarioRavasio:2024kem}
S.~Ferrario~Ravasio, R.~Gauld, B.~J{\"a}ger, A.~Karlberg, and G.~Zanderighi,
  ``{An event generator for neutrino-induced deep inelastic scattering and
  applications to neutrino astronomy},''
  \href{http://dx.doi.org/10.1140/epjc/s10052-025-14539-6}{{\em Eur. Phys. J.
  C} {\bf 85} (2025) no.~8, 888}, \href{http://arxiv.org/abs/2407.03894}{{\tt
  arXiv:2407.03894 [hep-ph]}}.

\bibitem{Bierlich:2022pfr}
C.~Bierlich {\em et al.}, ``{A comprehensive guide to the physics and usage of
  PYTHIA 8.3},'' \href{http://dx.doi.org/10.21468/SciPostPhysCodeb.8}{{\em
  SciPost Phys. Codeb.} {\bf 2022} (2022)  8},
  \href{http://arxiv.org/abs/2203.11601}{{\tt arXiv:2203.11601 [hep-ph]}}.

\bibitem{ParticleDataGroup:2018ovx}
{\bf Particle Data Group} Collaboration, M.~Tanabashi {\em et al.}, ``{Review
  of Particle Physics},''
  \href{http://dx.doi.org/10.1103/PhysRevD.98.030001}{{\em Phys. Rev. D} {\bf
  98} (2018) no.~3, 030001}.

\bibitem{faser_collaboration_2026_eg0mj-1ep53}
{\bf FASER} Collaboration, ``Measurement of muon neutrinos as a function of
  energy and rapidity with faser,'' Mar., 2026.

\bibitem{Godunov:2023myj}
S.~I. Godunov, E.~K. Karkaryan, V.~A. Novikov, A.~N. Rozanov, M.~I. Vysotsky,
  and E.~V. Zhemchugov, ``{Weak interaction corrections to muon pair production
  via the photon fusion at the LHC},''
  \href{http://dx.doi.org/10.1103/PhysRevD.108.093006}{{\em Phys. Rev. D} {\bf
  108} (2023) no.~9, 093006}, \href{http://arxiv.org/abs/2308.01169}{{\tt
  arXiv:2308.01169 [hep-ph]}}.

\bibitem{Francener:2024jra}
R.~Francener, V.~P. Goncalves, and D.~R. Gratieri, ``{Probing a $Z'$ gauge
  boson via neutrino trident scattering in the Forward Physics Facility at the
  LHC and FCC},'' \href{http://dx.doi.org/10.1140/epjc/s10052-025-14337-0}{{\em
  Eur. Phys. J. C} {\bf 85} (2025) no.~5, 601},
  \href{http://arxiv.org/abs/2411.04253}{{\tt arXiv:2411.04253 [hep-ph]}}.

\bibitem{Altmannshofer:2024hqd}
W.~Altmannshofer, T.~M{\"a}kel{\"a}, S.~Sarkar, S.~Trojanowski, K.~Xie, and
  B.~Zhou, ``{Discovering neutrino tridents at the Large Hadron Collider},''
  \href{http://dx.doi.org/10.1103/PhysRevD.110.072018}{{\em Phys. Rev. D} {\bf
  110} (2024) no.~7, 072018}, \href{http://arxiv.org/abs/2406.16803}{{\tt
  arXiv:2406.16803 [hep-ph]}}.

\bibitem{Francener:2026ukt}
R.~Francener and V.~P. Goncalves, ``{Neutrino trident process at a muon
  collider},'' \href{http://arxiv.org/abs/2610.00723}{{\tt arXiv:2610.00723
  [hep-ph]}}.

\bibitem{Kling:2026ghd}
F.~Kling, T.~M{\"a}kel{\"a}, S.~Trojanowski, and B.~Zhou, ``{Neutrino Tridents
  and $W$-boson production at the Muon Collider},''
  \href{http://arxiv.org/abs/2610.17263}{{\tt arXiv:2610.17263 [hep-ph]}}.

\bibitem{He:1990pn}
X.~G. He, G.~C. Joshi, H.~Lew, and R.~R. Volkas, ``{NEW Z-prime
  PHENOMENOLOGY},'' \href{http://dx.doi.org/10.1103/PhysRevD.43.R22}{{\em Phys.
  Rev. D} {\bf 43} (1991)  22--24}.

\bibitem{He:1991qd}
X.-G. He, G.~C. Joshi, H.~Lew, and R.~R. Volkas, ``{Simplest Z-prime model},''
  \href{http://dx.doi.org/10.1103/PhysRevD.44.2118}{{\em Phys. Rev. D} {\bf 44}
  (1991)  2118--2132}.

\bibitem{Buchmuller:1985jz}
W.~Buchmuller and D.~Wyler, ``{Effective Lagrangian Analysis of New
  Interactions and Flavor Conservation},''
  \href{http://dx.doi.org/10.1016/0550-3213(86)90262-2}{{\em Nucl. Phys. B}
  {\bf 268} (1986)  621--653}.

\bibitem{Boughezal:2020klp}
R.~Boughezal, C.-Y. Chen, F.~Petriello, and D.~Wiegand, ``{Four-lepton $Z$
  boson decay constraints on the standard model EFT},''
  \href{http://dx.doi.org/10.1103/PhysRevD.103.055015}{{\em Phys. Rev. D} {\bf
  103} (2021) no.~5, 055015}, \href{http://arxiv.org/abs/2010.06685}{{\tt
  arXiv:2010.06685 [hep-ph]}}.

\bibitem{Hencken:1993cf}
K.~Hencken, D.~Trautmann, and G.~Baur, ``{Electromagnetic electron - positron
  pair production in heavy ion collisions with impact parameter zero},''
  \href{http://dx.doi.org/10.1103/PhysRevA.49.1584}{{\em Phys. Rev. A} {\bf 49}
  (1994)  1584--1596}.

\end{thebibliography}\endgroup

\end{document}